\documentclass[]{spieman}
\usepackage[utf8]{inputenc}
\usepackage{graphicx}
\usepackage{hyperref}
\usepackage{subfig}
\usepackage{amsmath,amsfonts,amssymb}
\usepackage{setspace}
\usepackage[symbol]{footmisc}
\title{Long-Term Monitoring of Throughput in Las Cumbres Observatory's Fleet of Telescopes}
\author[a]{Daniel-Rolf Harbeck}
\author[a]{Curtis McCully}
\author[a]{Andrew Pickles}
\author[a]{Nikolaus Volgenau}
\author[a]{Patrick Conway}
\author[a]{Brook Taylor}
\affil[a] {Las Cumbres Observatory, Goleta, CA (USA)}

\begin{document}
\maketitle

\begin{abstract}
The Las Cumbres Observatory operates a fleet of robotically controlled telescopes currently two 2m,
nine 1m, and ten 0.4m telescopes, distributed amongst six sites covering both hemispheres.
Telescopes of an aperture class are equipped with an identical set of optical imagers, and those
data are subsequently processed by a common pipeline (BANZAI). The telescopes operate without direct
human supervision, and assessing the daily and long-term scientific productivity of the fleet of
telescopes and instruments poses an operational challenge.

One key operational metric of a telescope/instrument system is throughput. We present a method of
long-term performance monitoring based on nightly science observations: For every image taken in
matching filters and within the footprint of the PANSTARRS DR1 catalog we derive a photometric
zeropoint, which is a good proxy for system throughput. This dataset of over $250000$ data points
enables us to answer questions about general throughput degradation trends, and how individual
telescopes perform at the various sites. This particular metric is useful to plan the effort level
for on-site support and to prioritize the cleaning and re-aluminizing schedule of telescope optics
and mirrors respectively.
\end{abstract}

\keywords{Telescope operations, mirror reflectivity}

\section{The Las Cumbres Observatory Fleet of Telescopes and the Operational Challenge}

The Las Cumbres Observatory (LCOGT) operates a fleet of currently more than 20 telescopes,
consisting of ten 0.4 meter, nine 1 meter, and two 2 meter telescopes at multiple sites across the
globe\cite{brown2013}; an expansion of the network by additional five 1 meter class telescopes in
the northern hemisphere is planned. The telescopes are located at well-established observatory sites
which already provide service infrastructure\footnote{At LCOGT, telescope sites are identified by
    their nearest airport code (see Table \ref{tab_sites}), and we will follow that nomenclature    
    throughout this paper.}: Cerro Tololo Inter-American Observatory (CTIO) in Chile, South 
    African
Astronomical Observatory (SAAO) in Sutherland, Siding Spring observatory in Australia, Teide
Observatory in Canary Islands, Spain, McDonald Observatory in the continental US, and at Haleakala
Observatory on
Maui. Ensuring scientific productivity in this globally distributed observatory is a multi-faceted
challenge, where financial realities and the time overheads of international logistics need to be
considered.

The maintenance and support situation for LCOGT is complicated since all telescopes operate
autonomously, i.e., both nighttime and daytime operations of the network are based on computerized
execution of observations without direct local or remote human supervision. The observing schedule
for the entire observatory is dynamically generated in near real time to react to short term changes
in the telescope network\cite{saunders2014}, e.g., due to weather closures or a camera failure. The
sheer number of telescopes and the lack of human presence at the sites pose a challenge to
understand the state of a facility  and to discover problems of the telescope network. E.g., there
is no daily walk-through of the facility as is typically practiced at larger single telescope
facilities, and observatory staff's interaction with the telescopes fully depends on remote sensing.

With the exception of the two 2 meter telescope sites (Haleakala and Siding Spring) where LCOGT
staff are in residence, emergency maintenance and basic operational support are provided by the
observatory's site staff. The telescopes and site infrastructure are maintained and upgraded during
dedicated service campaigns. Since the effort and cost for each maintenance campaign to a site are
large --- equipment and personnel need to be transported from the headquarters in Santa Barbara, CA,
to each site --- it is highly desirable to prioritize service tasks based on performance metrics
that are relevant to the scientific productivity of the telescope network.

In this paper we describe how one specific metric, the throughput of a telescope system as measured
by the photometric zeropoint, is used to understand the state of the
telescope's optics and to guide improvement and maintenance throughout the observatory.

\begin{table}[h]
\centering
\begin{tabular} {|l|c|ccc|} \hline
Location                  & Site code & \#2m & \#1m & \#0.4m \\ \hline
Siding Spring, Australia & COJ & 1 & 2 & 2  \\
Sutherland, South Africa  & CPT &   & 3 & 1 \\
Teide, Spain              & TFN &   &   & 2 \\
Cerro Tololo, Chile       & LSC &   & 3 & 2 \\
McDonald Observatory, USA & ELP &   & 1 & 1\\
Haleakala, Hawaii, USA    & OGG & 1 &   & 2 \\\hline
                       &   \bf Total:  & \bf 2 & \bf 9 & \bf 10 \\ \hline
\end{tabular}
\vspace{1ex}
\caption{\label{tab_sites} List of observatories with LCOGT telescopes designed for science
operations. The LCOGT-internal site code is derived from the nearest major airport. }
\end{table}

\subsection{Maintenance of Optical Surfaces at LCOGT}

\begin{figure}
\includegraphics[width=0.95\textwidth]{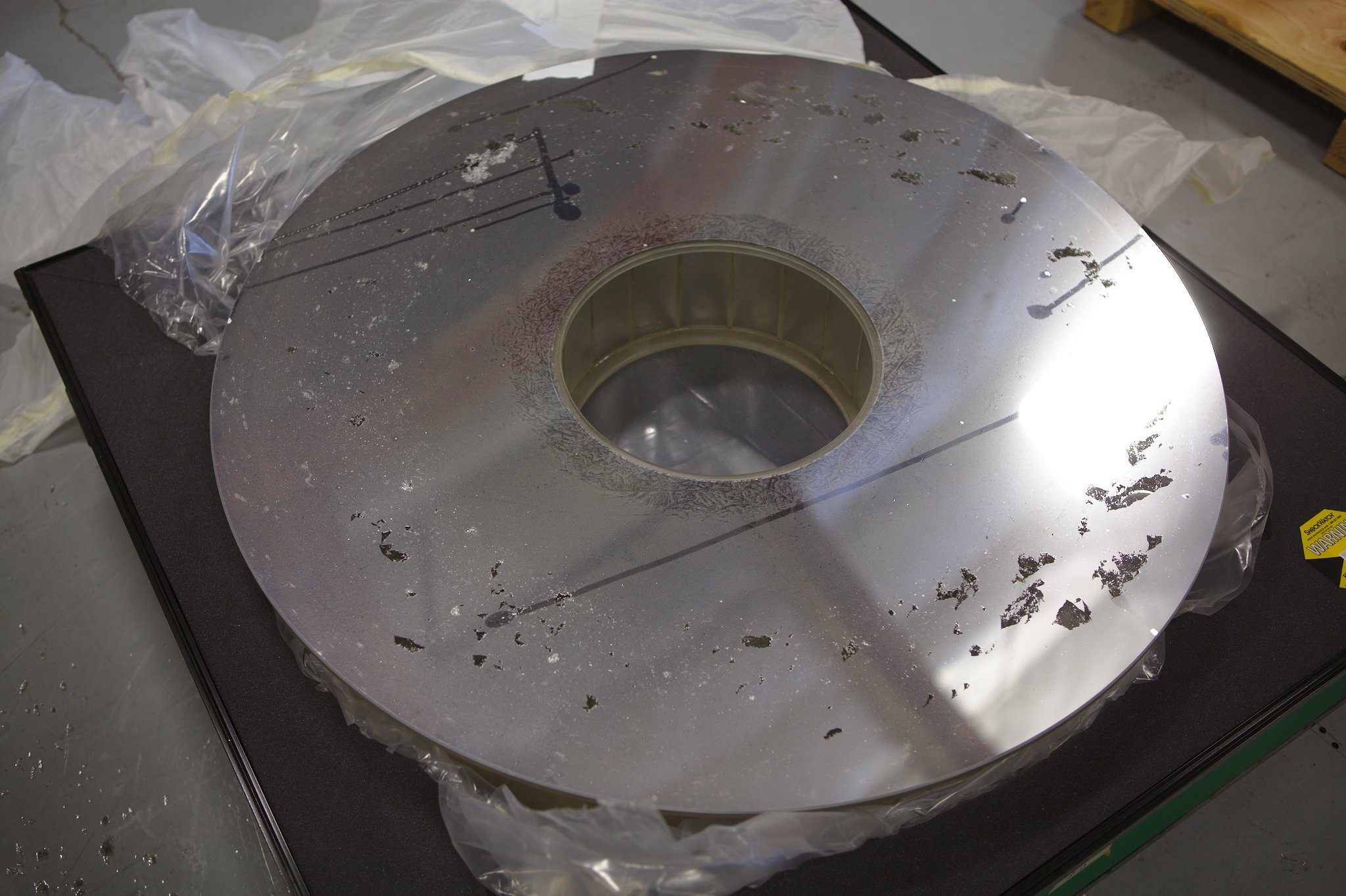}
\caption{\label{fig_mirrorpeel} Mirror returned form the LSC (CTIO) site after five years of
operation. The aluminum coating has peeled off, or is in the state of peeling of, at a significant
area of the mirrors. }
\end{figure}

The mirrors of a telescope are a key factor in the  scientific performance of an observatory. Their
reflectance typically degrades over time due to their exposure to the environment. A dominant
degradation mechanism is dust and pollen accumulation on the mirror surface which is usually
addressed by CO$_2$ snow cleaning or direct contact methods (e.g., wet wash with mild soap, adhesive
films). The long-term degradation of optics is driven by the accumulation of scratches, dust
permanently binding to the surface, and in worse cases, by the delamination of the aluminum coating
itself. The latter is a particular problem for all deployed LCOGT mirrors, where the aluminum layer
has peeled off and degraded over significant areas of the mirrors (Figure \ref{fig_mirrorpeel}).

The mirrors of the 1 meter telescopes still use the original coating from their deployment in 2012
and 2013, i.e., at the time of this writing they are of the order of five to six years old. The
coating  on the two meter telescope's  primary and secondary mirrors at the COJ site (Siding
Spring) are 5  and 16 years old, respectively. The primary  mirror at the OGG site (Haleakala) was
coated in 2017, after 8 years in service; the secondary  was coated in 2014 with a protective
overcoat.

Site staff administers CO$_2$-snow cleaning of the mirror surfaces of the 1 meter and 2 meter
telescopes on a roughly monthly basis, following locally adopted best practices. However, there is
no widespread cleaning procedure established for the 0.4 meter's corrector plates yet, although
they accumulate dust as well. LCOGT is working currently on improving the cleaning protocol for all
telescopes. In order to address the significantly degraded coating at the 1 meter and 2 meter
telescopes, all mirrors are scheduled for recoating by the end of 2018. This is an enormous
logistical effort given there are eleven such mirrors at multiple sties.

At the beginning of 2018 a project team was formed to complete the optics renewal for all 1 meter and
2 meter telescopes.  Leveraging a quality-controlled coating company, a schedule was outlined to complete
the removal, reassembly, and collimation of the network of telescopes.  Revised 
existing procedures were tested at LCOGT's headquarter, where personnel were trained to complete the
work in the field.  Collimation characterization criteria were developed to qualify the newly
installed optics along with wavefront sensing equipment to validate the before and after condition
of the telescopes.  Upon completion of the ELP (McDonald Observatory) 1 meter telescope, the team 
is now formalizing the maintenance tasks for cleaning and monitoring the throughput of the network's 
telescopes.

\section{Measuring the photometric zeropoint in a homogeneous fleet of telescopes}

Our interest is in using the photometric zeropoint measured in a science image as a proxy for the
telescope and image system throughput, as it is standard practice at many observatories. An
individual image's photometric zeropoint measurement is impacted by many factors, such as
attenuation by clouds, measurement errors, and instrument errors, and is hence of limited relevance.
However, the long term trend in the photometric zeropoint of all images taken at a telescope is a
meaningful proxy for the system's efficiency.

On-sky throughput measurements in the past have often been limited to isolated observations of a few
standard stars per year\cite{benn2000}, or to nightly standard star observations\cite{hopp2008}. We
exploit the availability of well-calibrated photometric catalogs and access to a central
data archive with homogeneously calibrated data to turn {\em every} exposure at an LCOGT telescope
into a measurement of a telescope's throughput. We have demonstrated this approach before for data
from the One Degree Imager at the WIYN 3.5 meter telescope\cite{harbeck2014}.

Each telescope is equipped with a primary scientific imaging camera (the imagers at the 1 and 2
meter telescopes are based on Fairchild CCD486 devices with a field of view of $27' \times 27'$ and
$10' \times 10'$ respectively, the imagers at the 0.4 meter telescopes are commercial SBIG 6303
cameras with a field of view of $30' \times 20'$). The cameras have a comprehensive filter set,
which includes a set in the Sloan Digital Sky Survey (SDSS) photometric system\cite{fukugita1996}.
The 2 meter telescopes provide low resolution spectroscopy capability via the
Floyds\cite{brown2013} spectrograph, whereas three of LCOGT's 1 meter telescopes are equipped with
high resolution Echelle spectrograph instruments (NRES)\cite{eastman2014,siverd2016}.

All data taken at LCOGT's telescopes are transferred via the Internet to a live storage located at
the headquarters in Santa Barbara. The incoming imaging data are then processed daily with the BANZAI
pipeline (Beautiful Algorithms to Normalize Zillions of Astronomical Images)
\cite{mccully2018}\footnote{\url{https://github.com/LCOGT/banzai}}, which  performs the standard
CCD processing (crosstalk, overscan, bias, dark, and flat field correction), followed by the
extraction of an object catalog from the image and cross-matching with existing catalogs to derive an
astrometric solution via {\tt astrometry.net}\cite{lang2010}. The object catalog is stored along
with the final calibrated data product and then published in LCOGT's online
archive\footnote{\url{https://archive.lco.global}}. Photometric calibration of images is not a part
of standard processing.

To measure the photometric zeropoint in LCOGT's images we have developed a program that crawls
through BANZAI processed images from LCOGT's on-site archive and derives a photometric zeropoint
calibration for each image by cross-matching sources with the PANSTARRS DR1 photometric
catalog\cite{chambers2016}. For each BANZAI-processed image taken at any telescope, we determine the
photometric zero point following the prescription below. Note that this approach is passive in the
sense that it is utilizing science observations that have already been observed -- no special
calibration field observation requests are injected into the schedule for the purpose of this
study, therefore maximizing usable science observations:

\begin{enumerate} 
    \item Check if an exposure is viable for photometric zeropoint measurement:
   
    \begin{itemize} 
        \item Image is taken in a filter with a viable transformation to the SDSS
        photometric system. At this time, the analysis is limited to the g' r' i' z' filters. 
        
        \item Image is in footprint of the PANSTARRS DR1 catalog, i.e., is the declination larger
         than $-30^\circ$ . 
        
        \item A minimum exposure time requirement is met: This is selected to be 10 seconds for the
         1 and 2 meter telescopes, and 15 seconds for the 0.4 meter telescopes. 
        
    \end{itemize}
    
    \item Cross-match the image's object catalog by RA/Dec coordinates with the PS1 catalog (a 
    locally cached copy is used). 
    
    \item Transform the PS1 photometry into the SDSS system\cite{finkbeiner2016}.
    
    \item Iteratively, with outlier rejection, fit the photometric zeropoint between instrumental
    magnitudes and catalog. \item Iteratively, with outlier rejection, fit a linear color term for
    diagnostic purposes: $$ {\rm mag}_{\rm SDSS} - {\rm mag}_{\rm inst} = {\rm zeropoint} + c \times
    ({\rm g'}-{ \rm r'})_{\rm SDSS}$$ The color term is not yet incorporated into the zeropoint
    fitting. \item Store the resulting zeropoint along with meta data (exposure identifier, 
    telescope, instrument identifier, airmass, formal error, color term) in a SQLite database. 
\end{enumerate}

\begin{figure}
\centering
\includegraphics[width=0.49\textwidth]{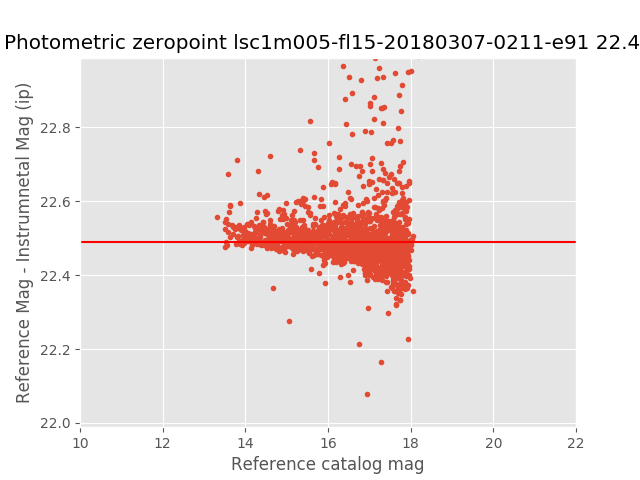} \hspace*{\fill}
\includegraphics[width=0.49\textwidth]{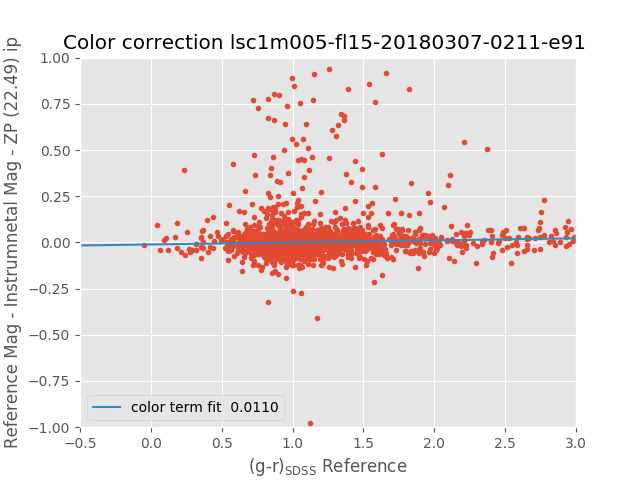} \\[1ex]
\caption{\label{fig_singleimageexample}Example of photometric zeropoint fit (left), and measurement of
airmass term for a single image exposure.}
\end{figure}

Diagnostic plots are optionally generated during the zeropoint fitting procedure to display the
SDSS reference band magnitude versus difference in instrumental and catalog magnitude; these plots
have been useful during code development, but they are too numerous to inspect for all data in the
archive and therefore disabled by default. We find those plots however useful when recommissioning a
camera at a telescope. Example plots are shown in Figure \ref{fig_singleimageexample}.

After initially processing all BANZAI-calibrated\footnote{BANZAI was deployed early 2016, which 
is why the following plots will start at that time; LCOGT has been operational earlier.} images 
in the LCOGT archive (over $250,000$ images), zeropoints for newly observed  images are now 
calculated on a daily basis using an autonomous process. The results  
are made available at a web page internal to LCOGT. It is a goal to implement the zeropoint 
calibration as a part of the default processing in the BANZAI pipeline.

\section{Long term trends in throughput}

The photometric zeropoint database by now contains over $250,000$ data points, spawning a parameter
space of about 30 cameras at 21 telescopes in four filters (g'r'i'z') from 2016 to today. A
canonical way to inspect those data is to show the photometric zeropoint of the imager (there is
only one primary imager at any moment per telescope) vs time, and we plot the zeropoint in the SDSS
g' and r' filter for all LCOGT's sites in Figure \ref{fig_zpLSC}; the sampling in the i' and z'
band filters is too sparse to add meaningful information. Different cameras are identified in those
plots. Photometric zeropoints are corrected to an airmass of 1 for plotting and further calculation.
Note that the zeropoint is in magnitudes, i.e., it is a logarithmic representation of the
throughput. A zeropoint difference $\Delta m$ translates into a flux (throughput) ratio as $I_1 /
I_2 = 10^{-0.4 \times \Delta m}$.  We observe:

\begin{enumerate}

\item Only the upper envelope of the overall zeropoint vs time trend is of importance - data points
below the mean trend line are an indication of non-photometric conditions. Measurements with large
errors (e.g., images taken in the Galactic plane) are grayed out in the plots. Some 0.4 meter 
telescopes do not yet have a long history of observations since they were only deployed recently.

\item Different CCD camera types have a different quantum efficiency, as seen by, e.g., the sudden
increase of the throughput in October 2016 in the throughput of the LSC doma-1m0a plot. At that time
an older SBIG CCD camera was replaced by a more sensitive Sinistro imager.

\item The general trend in zeropoint vs time trends down, as is expected.

\item Realuminizing of a mirror results in very significant throughput improvements. Those events 
(as well as a special cleaning of a mirror) are marked with grey vertical bars in the plots:
 \begin{enumerate}
 	\item In October 2017 the 1 meter telescope in the dome C at the LSC (CTIO) site.
    \item In November 2017 for the 2 meter telescope at the OGG (Haleakala) site. 
    \item In April 2018 the 1 meter telescope at the ELP  (McDonald observatory) site.
\end{enumerate}
 The throughput improved by about 0.8 magnitudes for the two one meter telescope, which is about a 
 factor of 2.1. The throughput of the 2 meter telescope improved by about 1.7 magnitudes, or a 
 factor of 4.8!

\item Cleaning a mirror, or for the 0.4 meter telescopes, the  corrector plate, can yield
a significant throughput improvement. An example is the 0.4 meter telescope at the COJ (Siding 
Spring) site around July 2017, where after CO$_2$ snow cleaning the throughput increased by about 
0.35 magnitudes, or a factor of 1.4.

\end{enumerate}

\begin{figure}
\centering 
LSC \\ 
\rule{\textwidth}{0.4pt} \\
\includegraphics[width=0.49\textwidth]{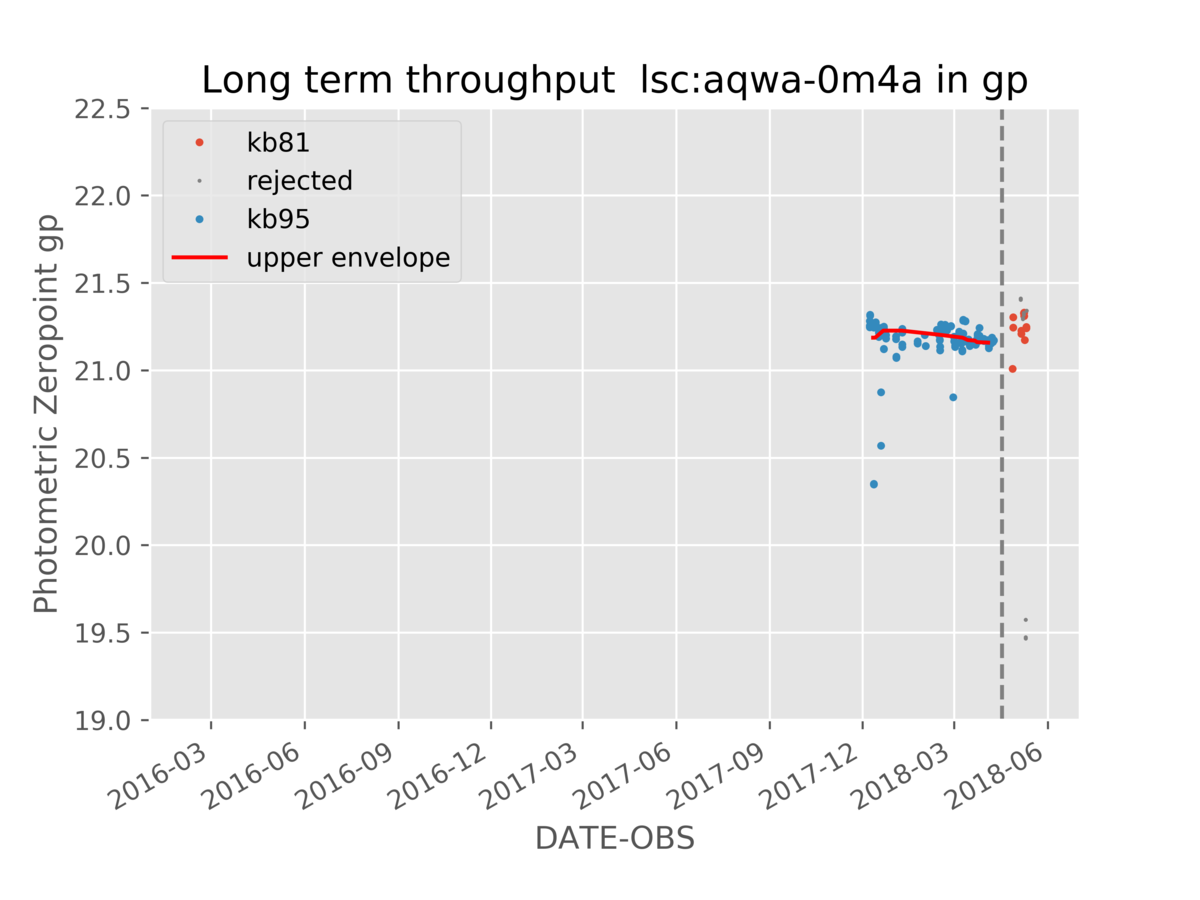} \hspace*{\fill}
\includegraphics[width=0.49\textwidth]{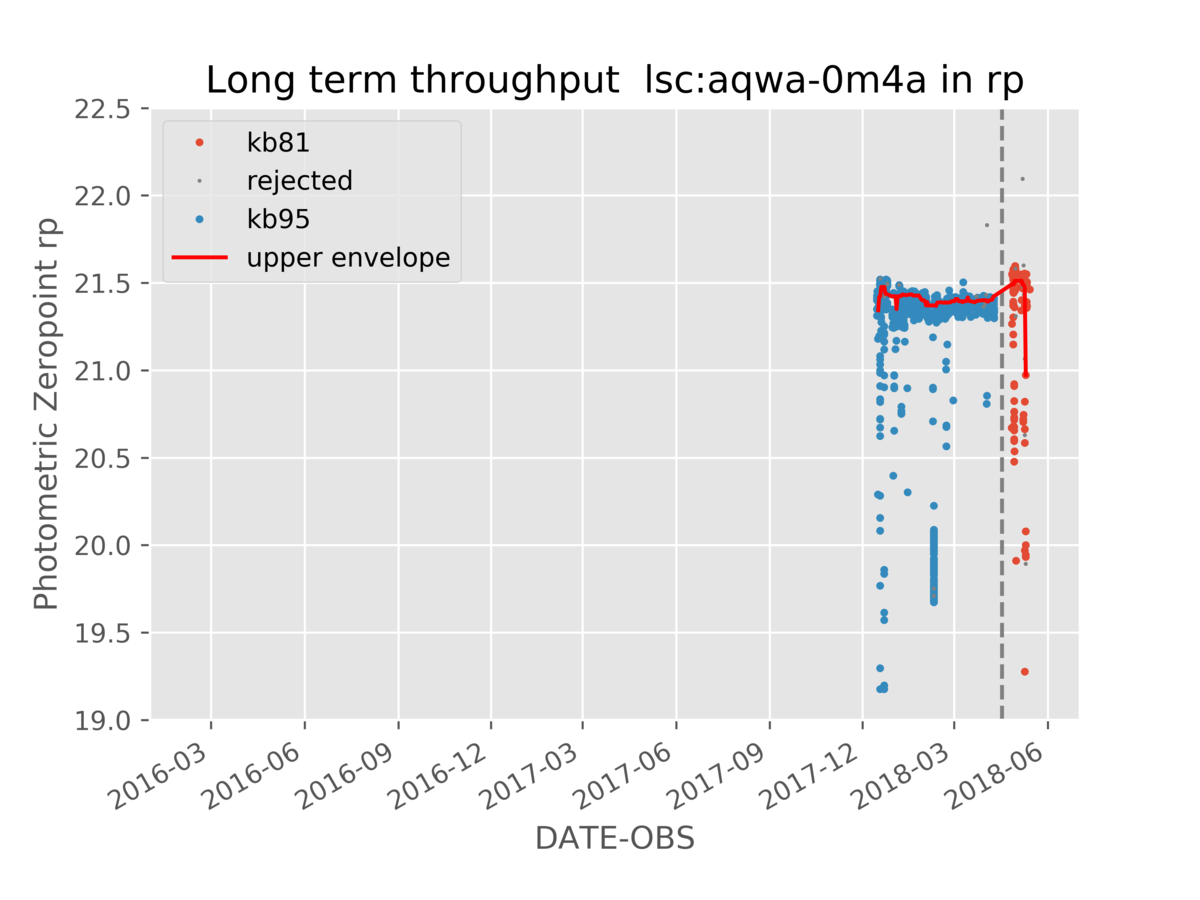} \\
\includegraphics[width=0.49\textwidth]{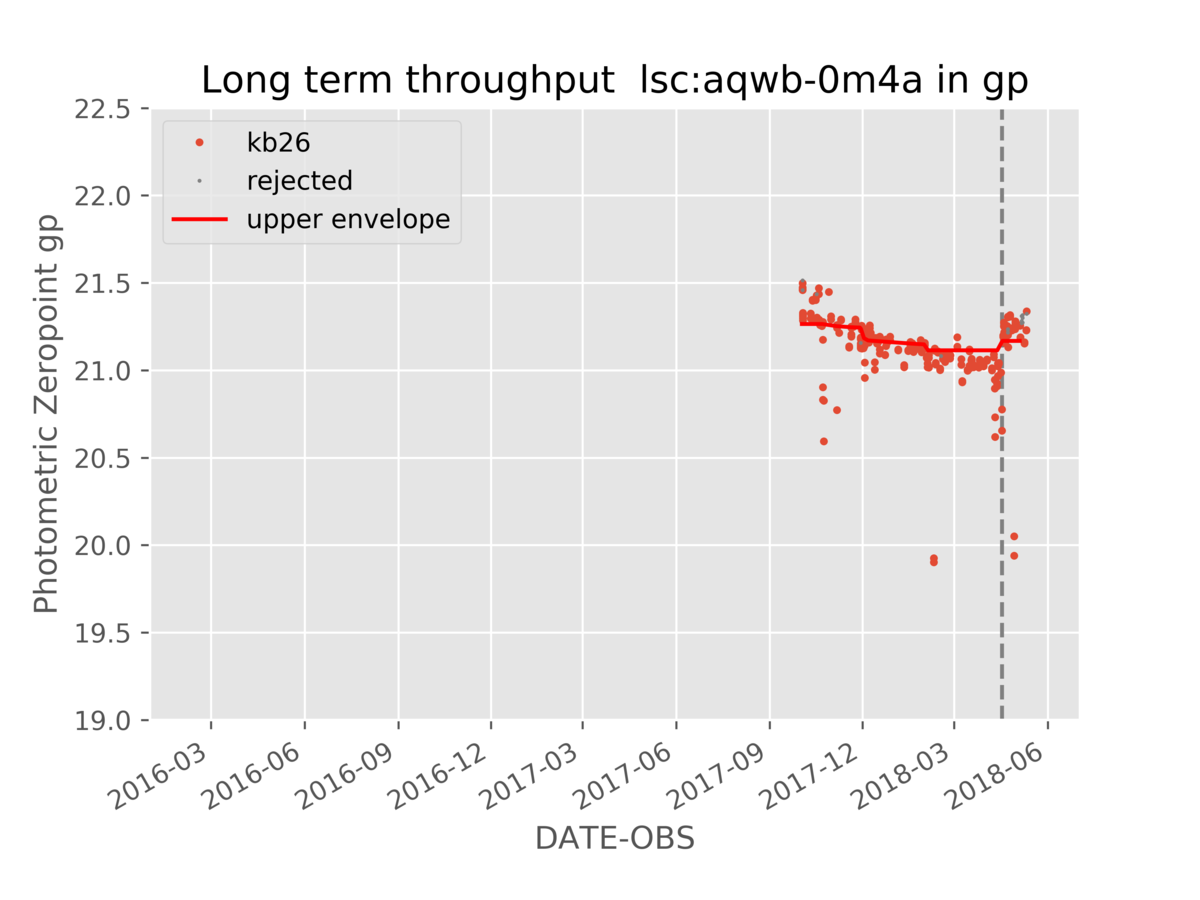} \hspace*{\fill}
\includegraphics[width=0.49\textwidth]{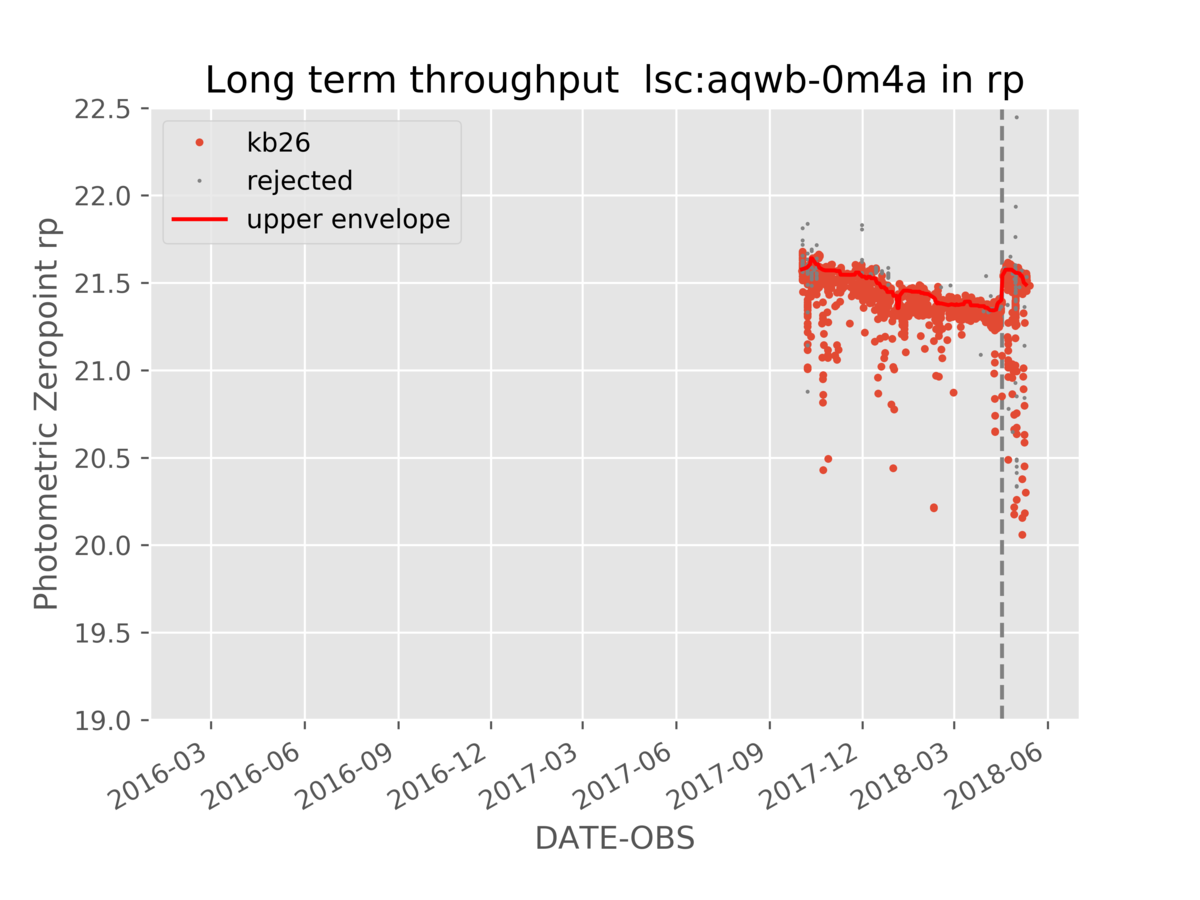} \\
\includegraphics[width=0.49\textwidth]{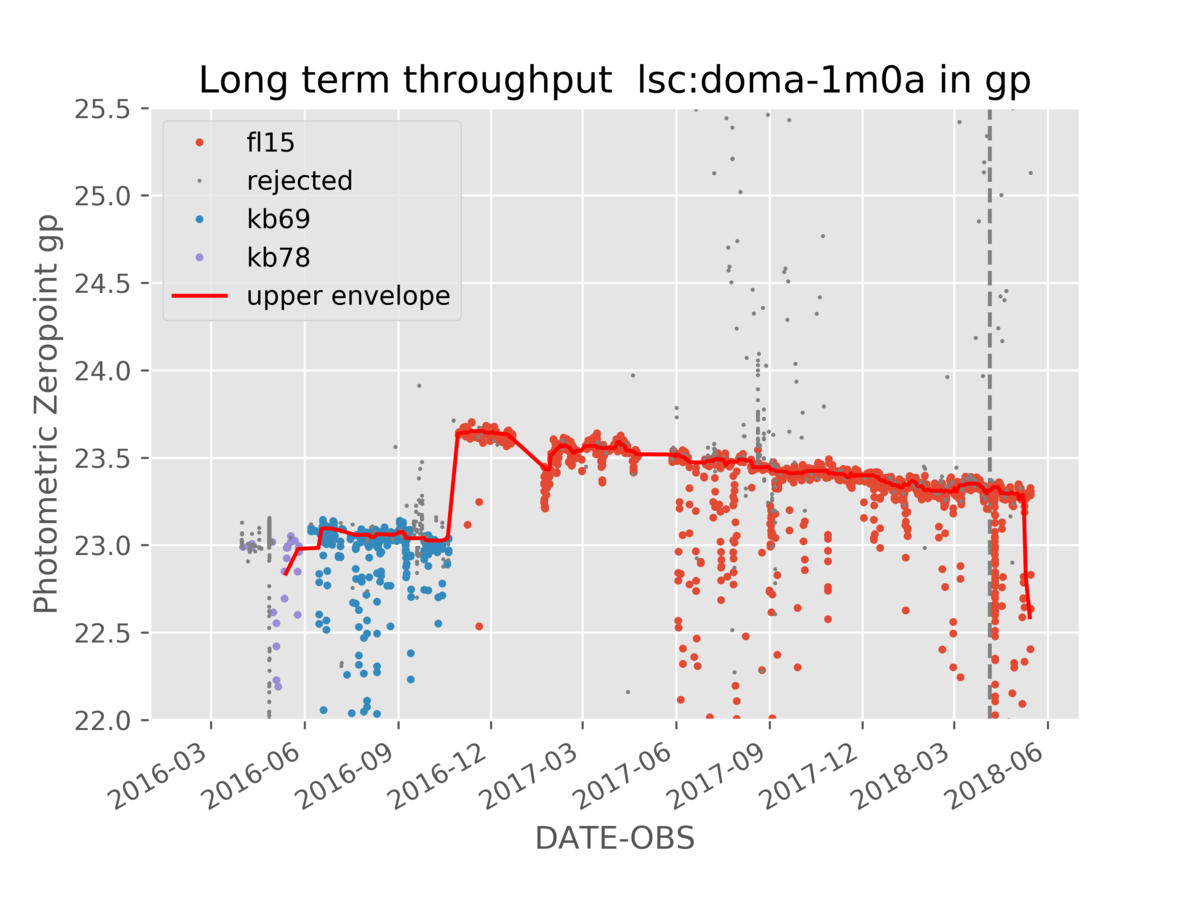}  \hspace*{\fill}
\includegraphics[width=0.49\textwidth]{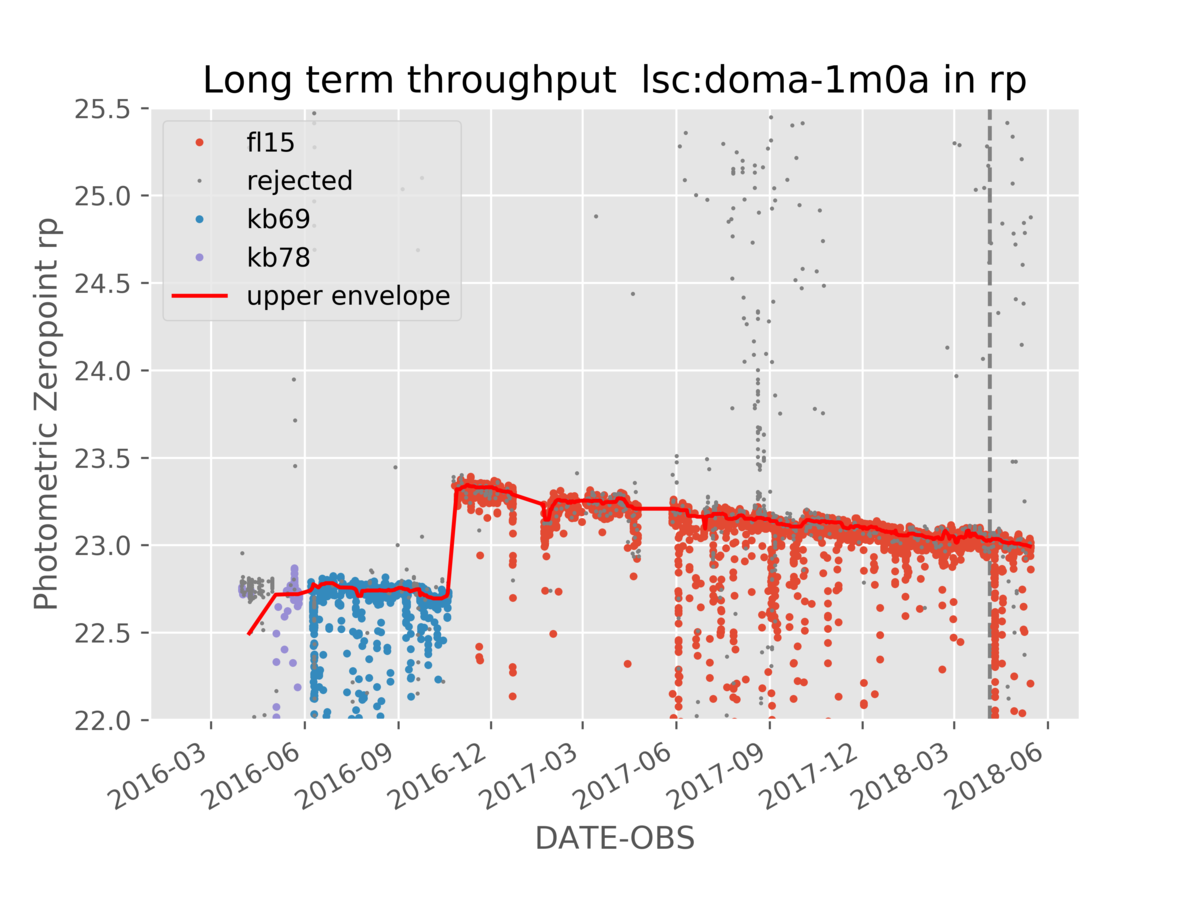} \\[1ex]

\caption{\label{fig_zpLSC} Photometric zeropoint trends for all LCOGT telescopes in science
operations. For each telescope we plot the air-mass corrected zeropoint in the SDSS g' and r' band
filters vs. date. Telescopes are identified by their site code, their enclosure, and telescope
identifier (e.g., lsc:doma-1m0a is the first 1 meter telescope in doma at lsc (CTIO)). Each imaging
camera at a telescope is identified by different colored dots. The red line indicates a fit of the
zeropoint over time. Event where the mirrors were realuminized (and some selected cleaning events)
are indicated by a vertical dashed grey line.}

\end{figure}

\begin{figure}\ContinuedFloat
\centering 
LSC continued \\ 
\rule{\textwidth}{0.4pt} \\
\includegraphics[width=0.49\textwidth]{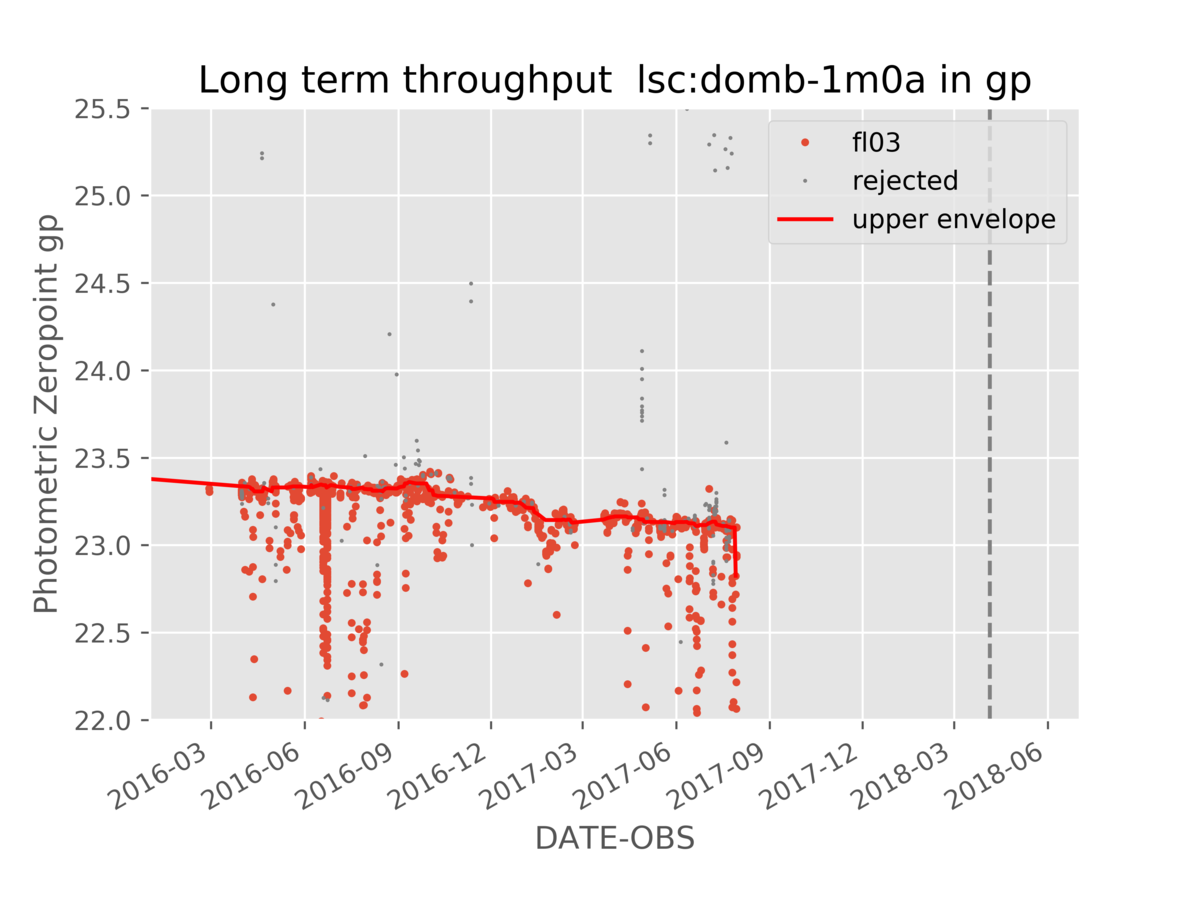} \hspace*{\fill}
\includegraphics[width=0.49\textwidth]{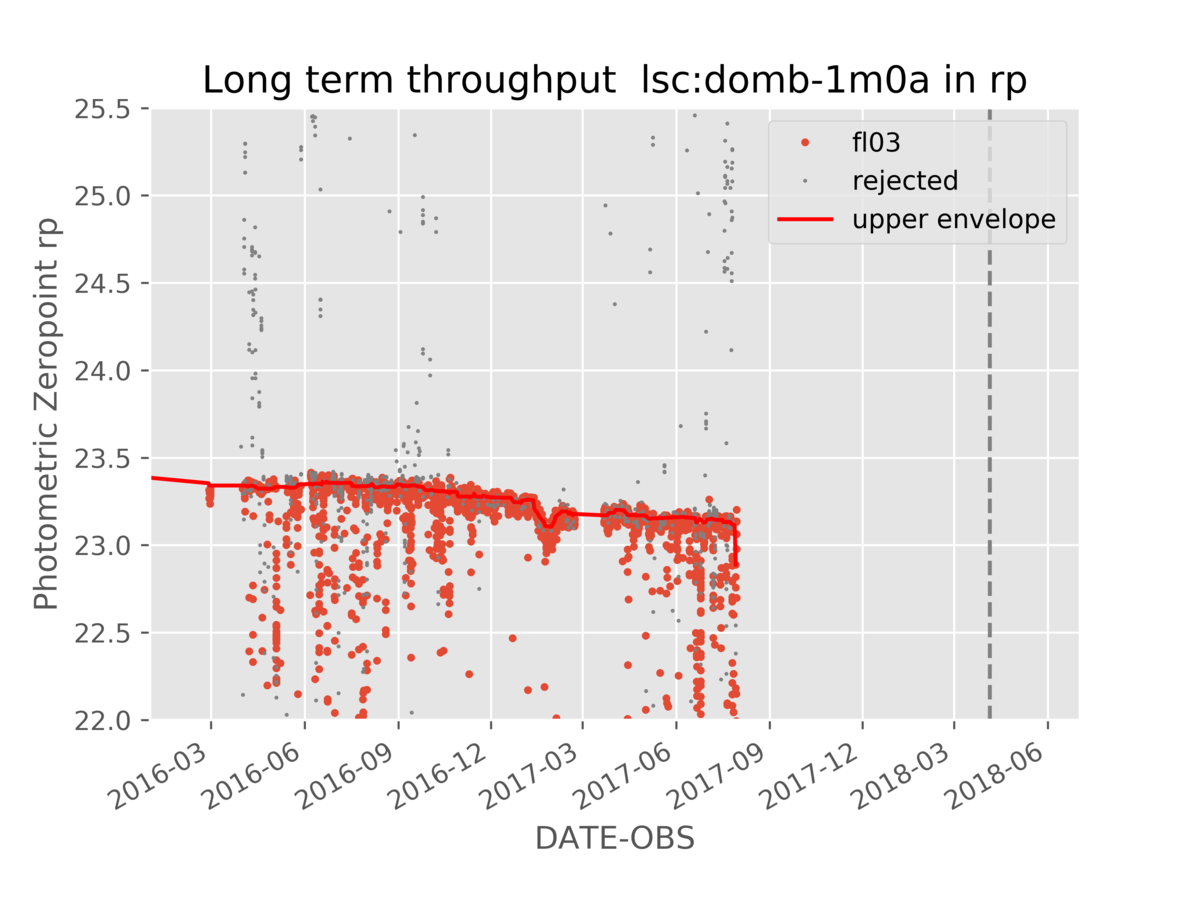} \\
\includegraphics[width=0.49\textwidth]{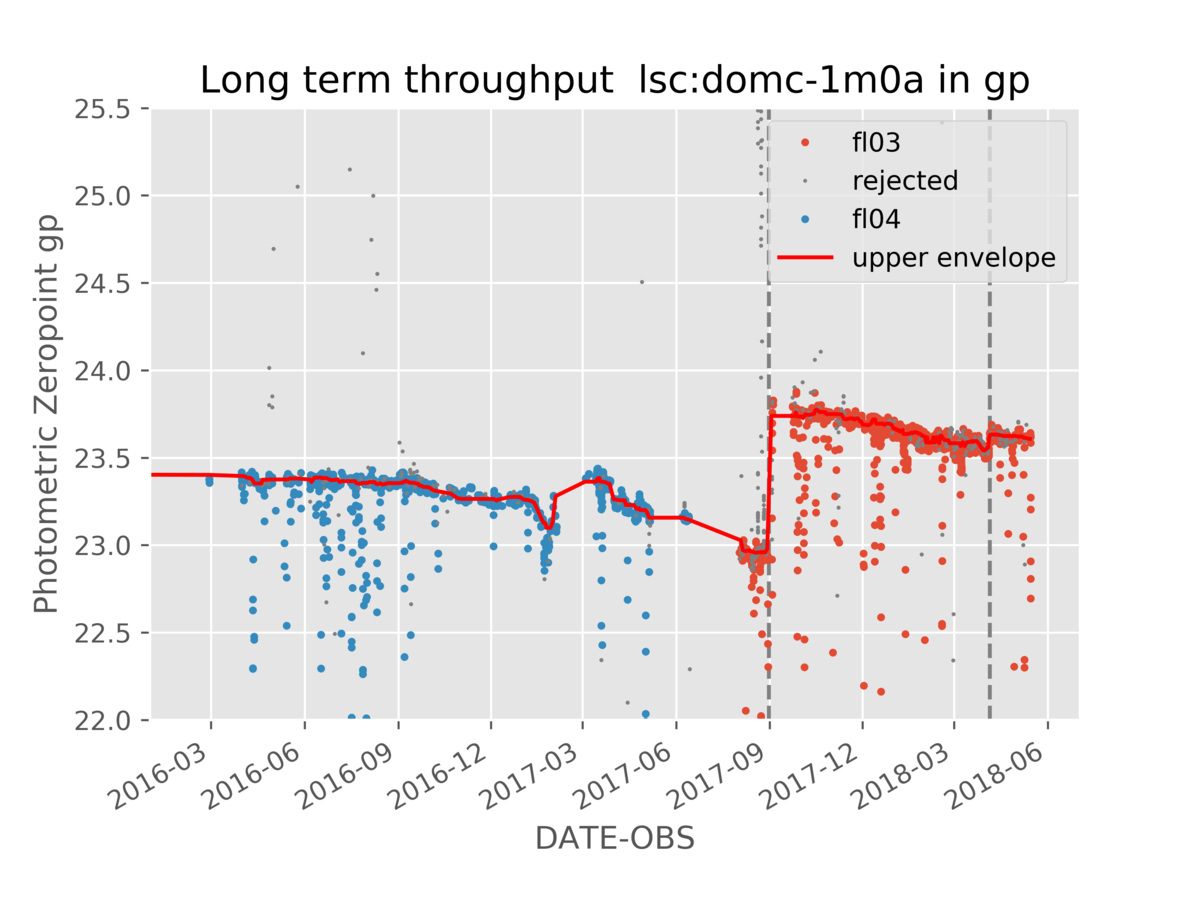} \hspace*{\fill}
\includegraphics[width=0.49\textwidth]{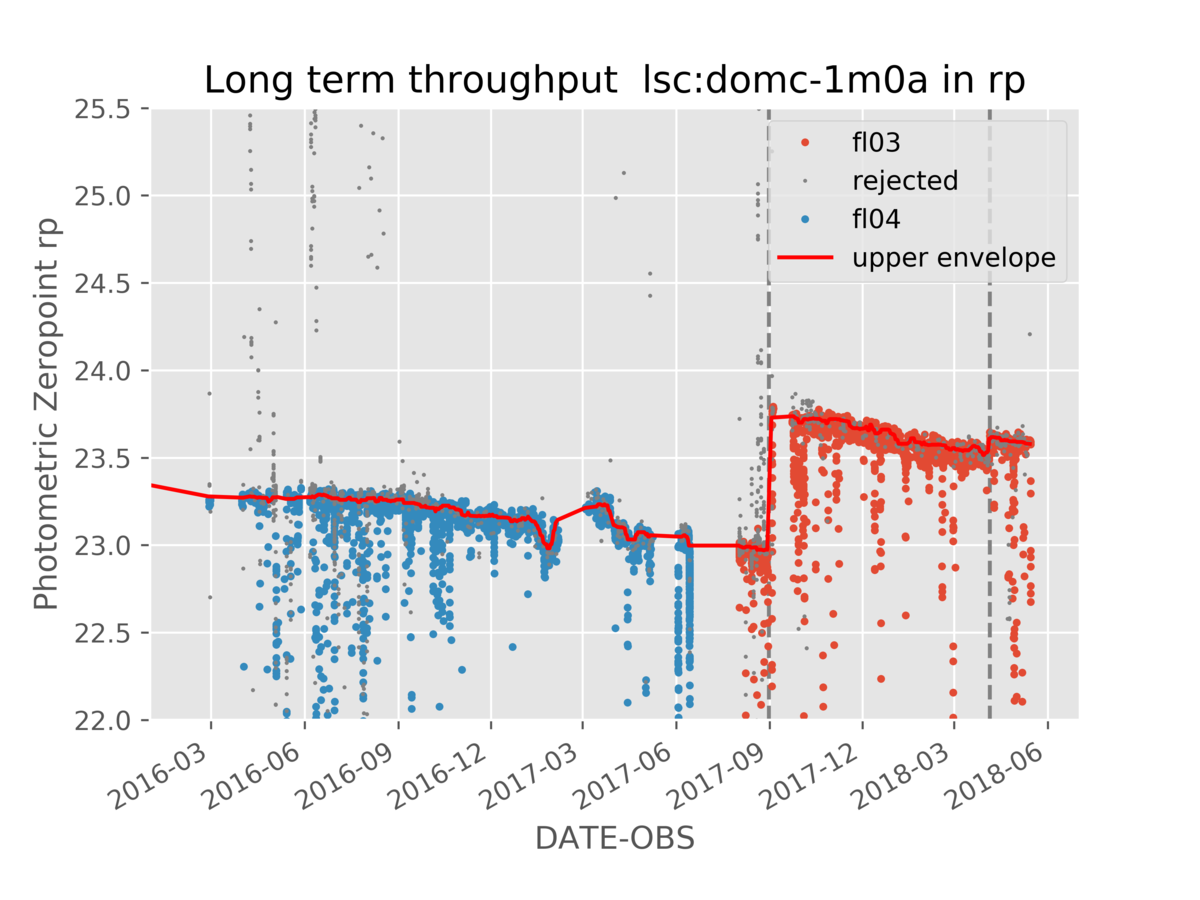} \\[1ex]
COJ \\ 
\rule{\textwidth}{0.4pt} \\
\includegraphics[width=0.49\textwidth]{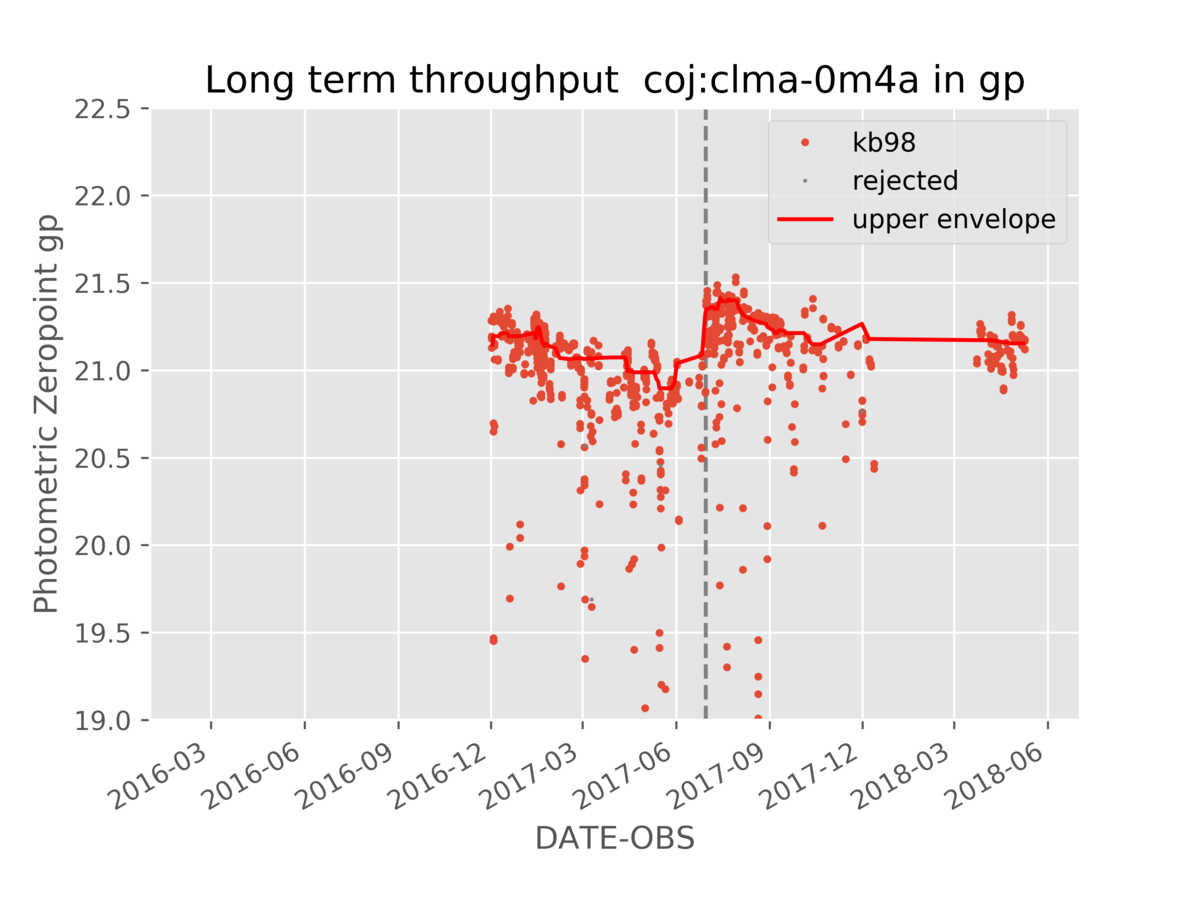} \hspace*{\fill}
\includegraphics[width=0.49\textwidth]{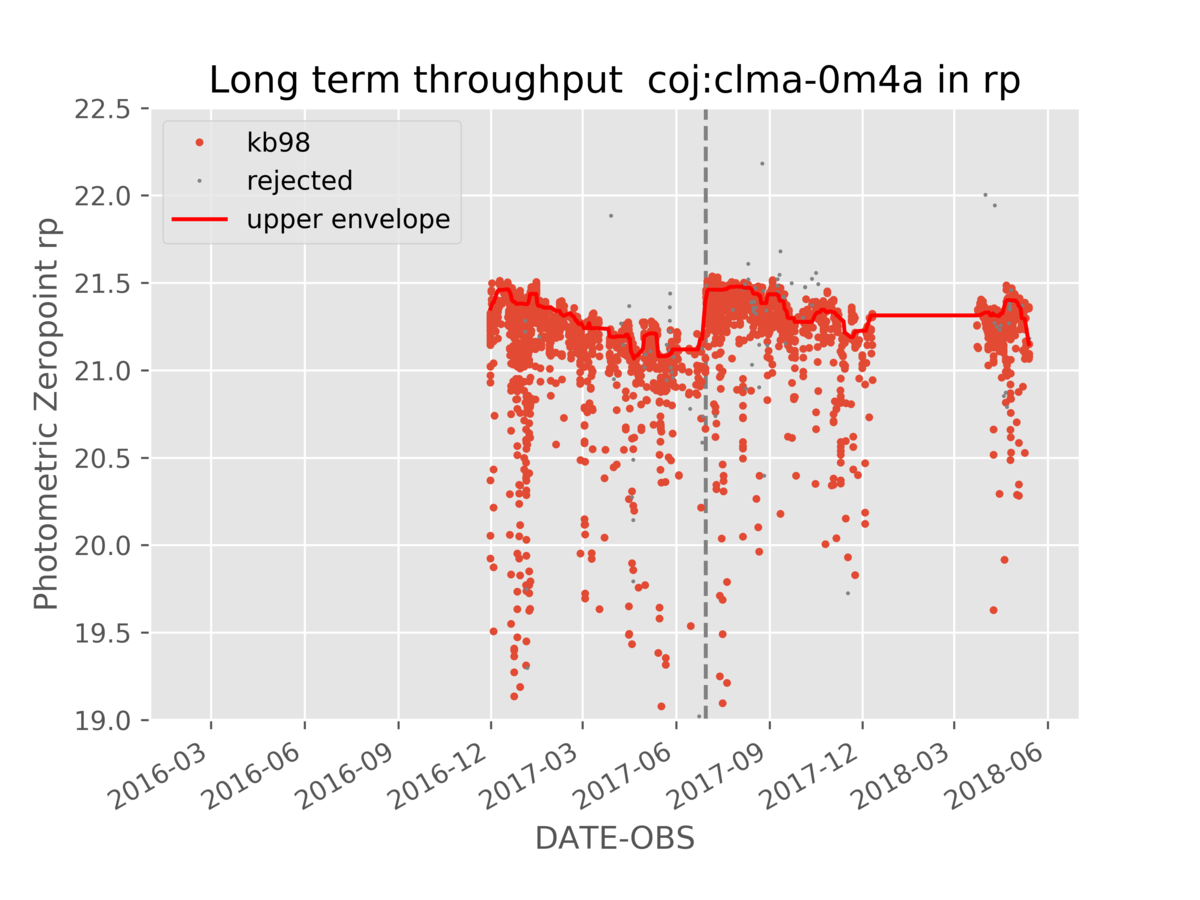} \\[1ex]
\caption {continued}
\end{figure}

\begin{figure}\ContinuedFloat
\centering 
COJ continued \\ 
\rule{\textwidth}{0.4pt} \\
\includegraphics[width=0.49\textwidth]{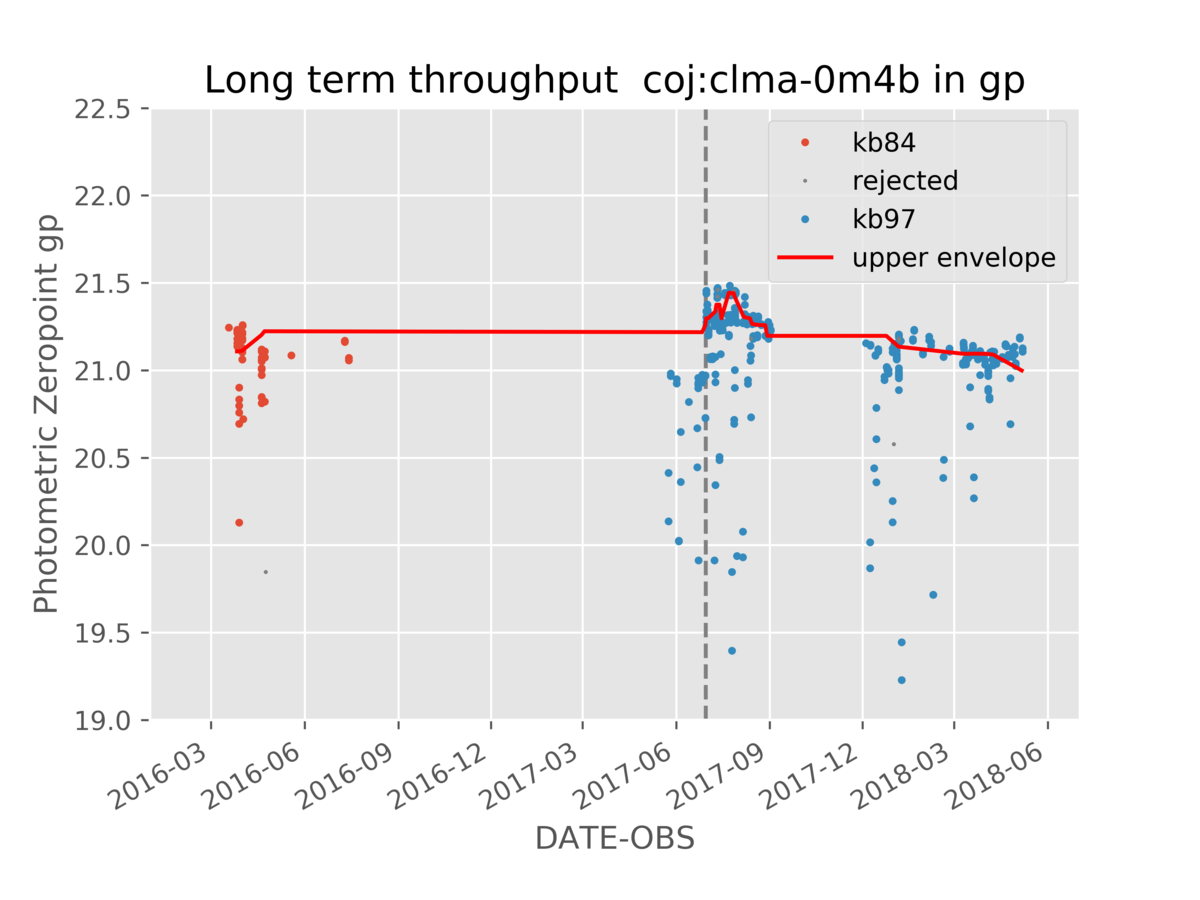} \hspace*{\fill}
\includegraphics[width=0.49\textwidth]{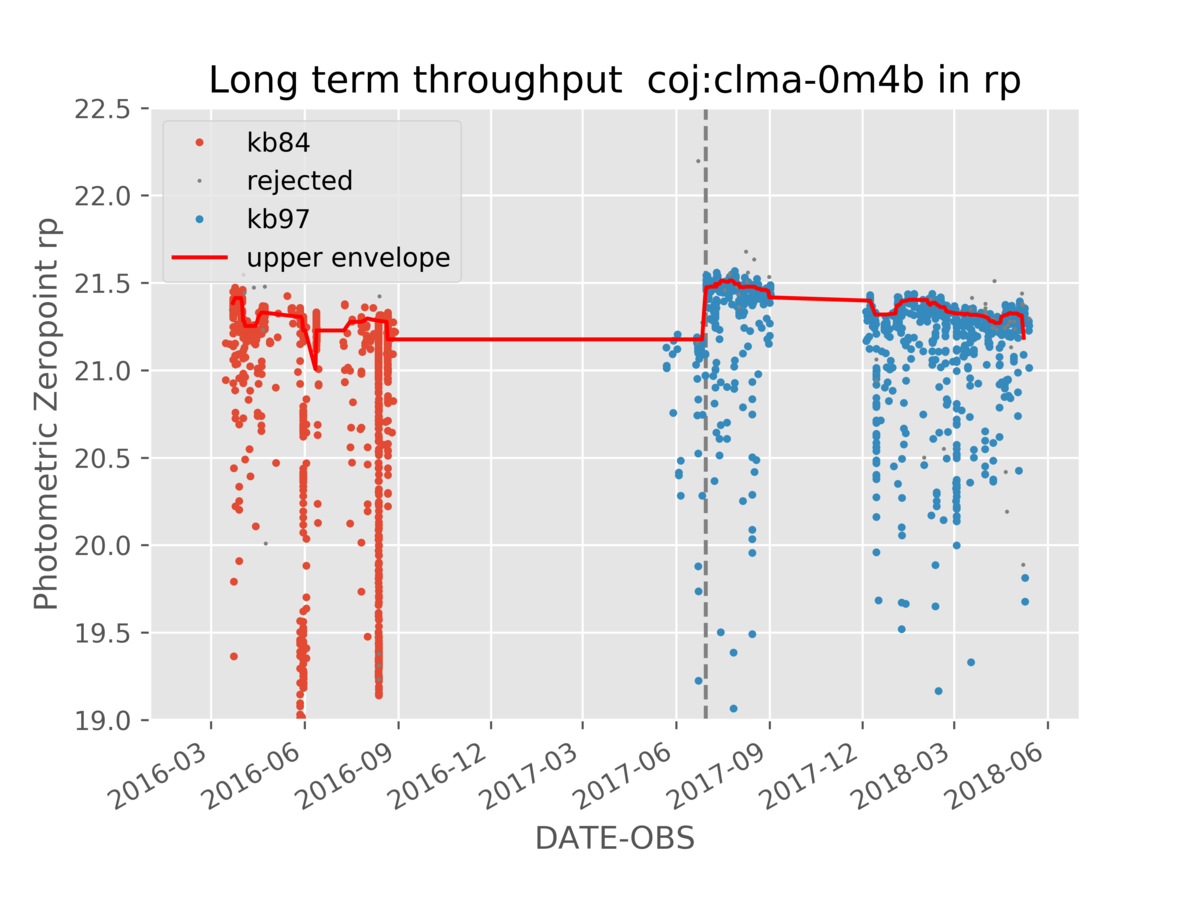} \\
\includegraphics[width=0.49\textwidth]{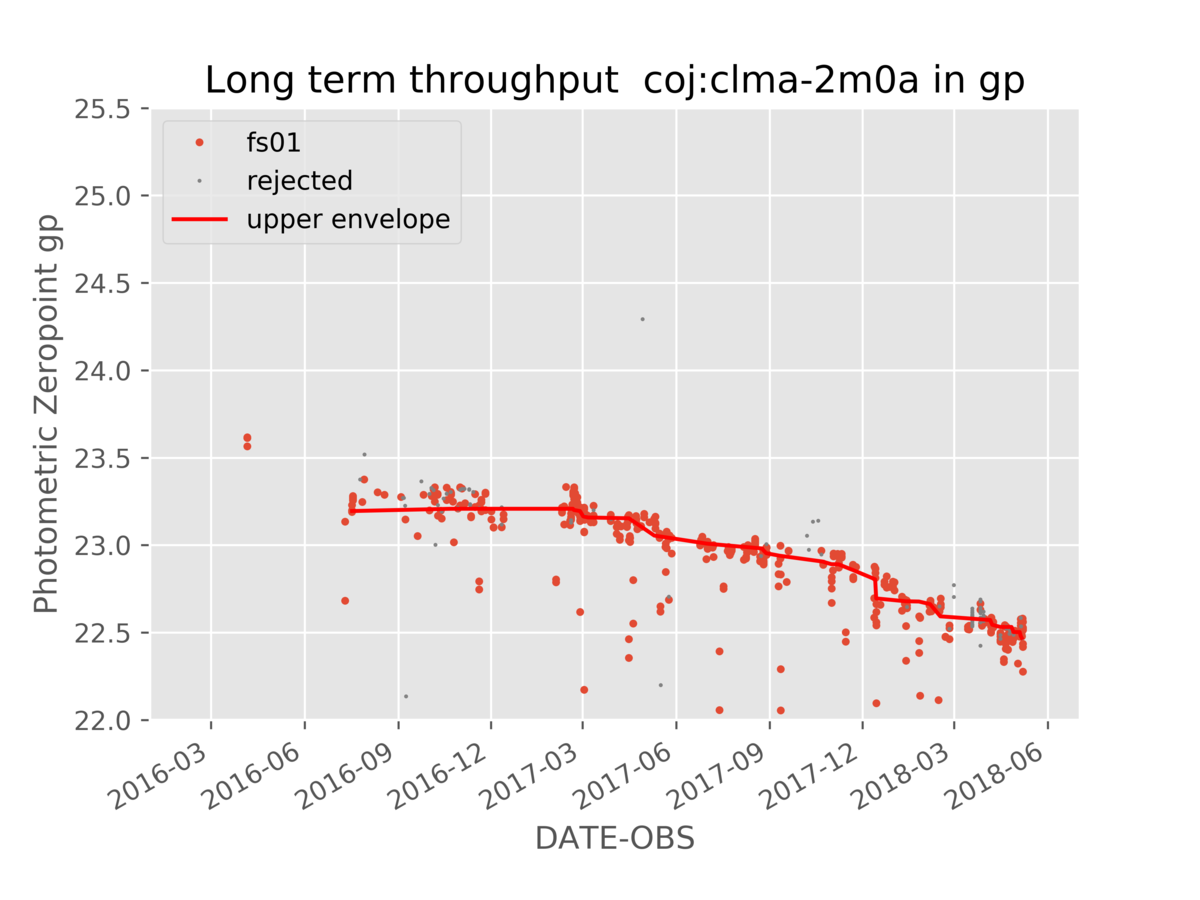} \hspace*{\fill}
\includegraphics[width=0.49\textwidth]{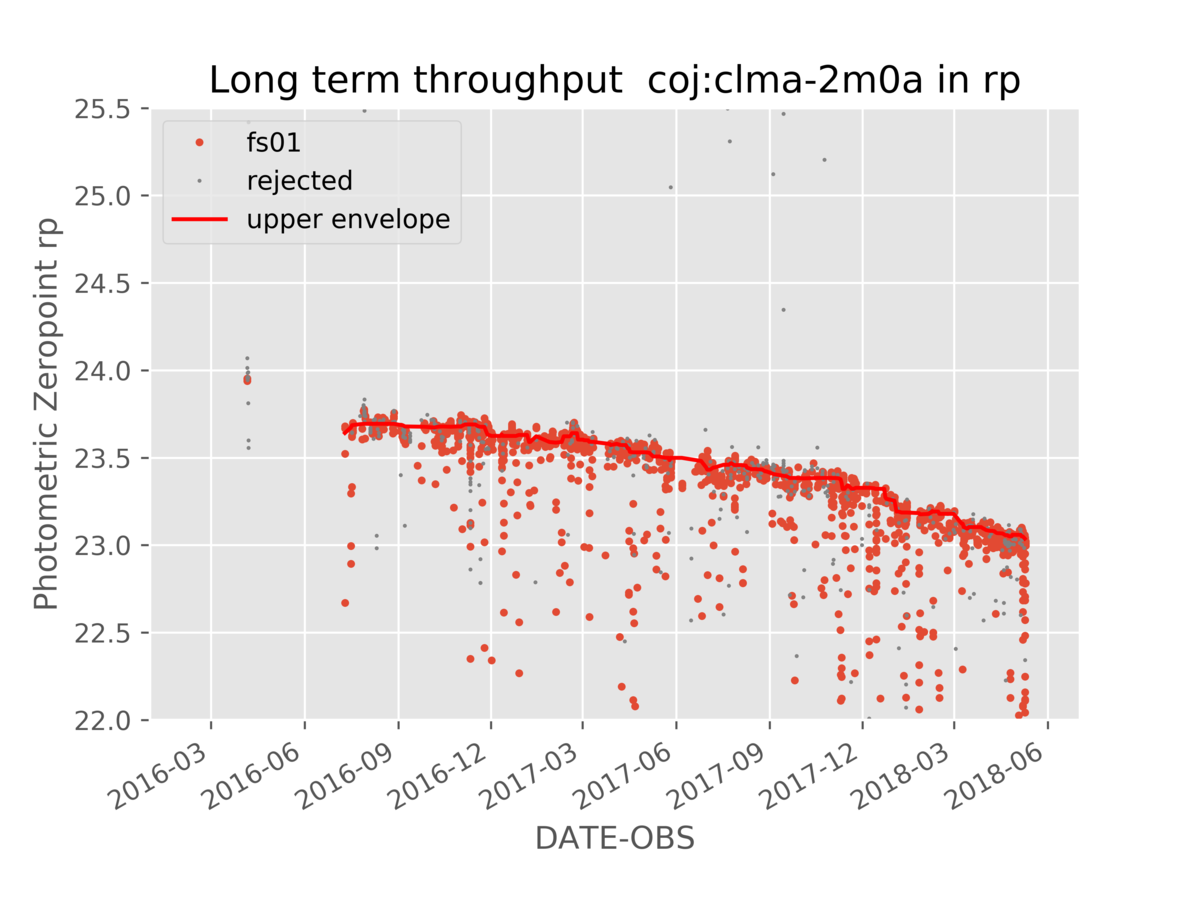} \\ 
\includegraphics[width=0.49\textwidth]{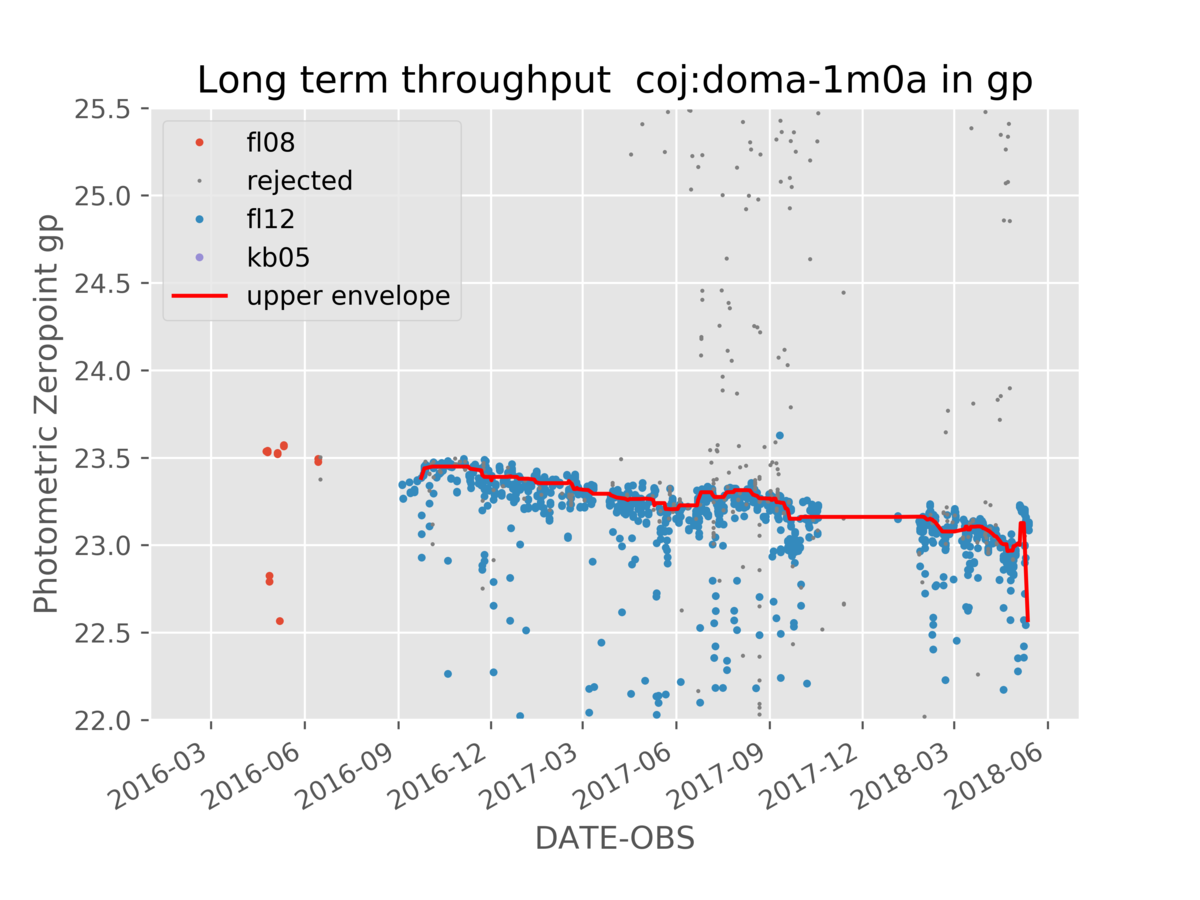} \hspace*{\fill}
\includegraphics[width=0.49\textwidth]{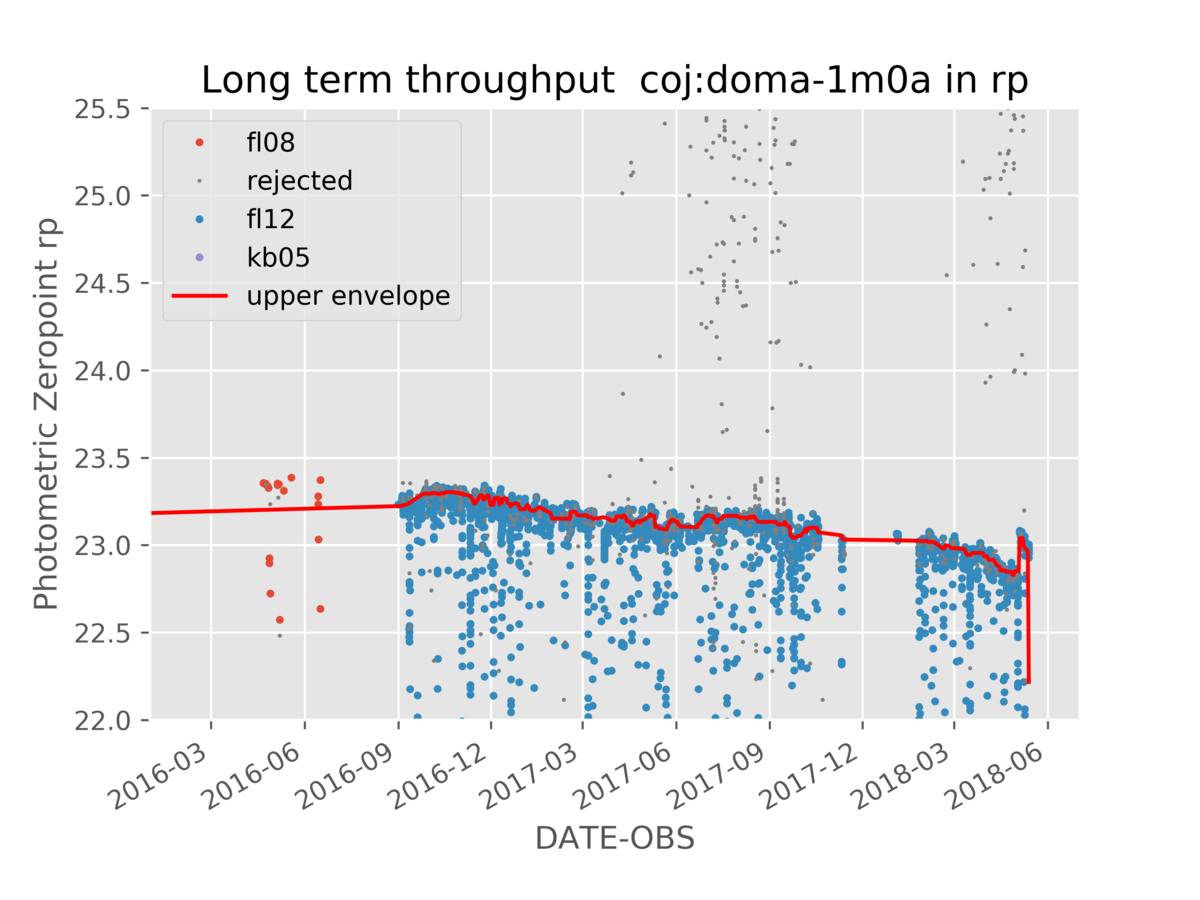} \\[1ex]
\caption {continued}
\end{figure}

\begin{figure}\ContinuedFloat
\centering 
COJ continued \\ 
\rule{\textwidth}{0.4pt} \\

\includegraphics[width=0.49\textwidth]{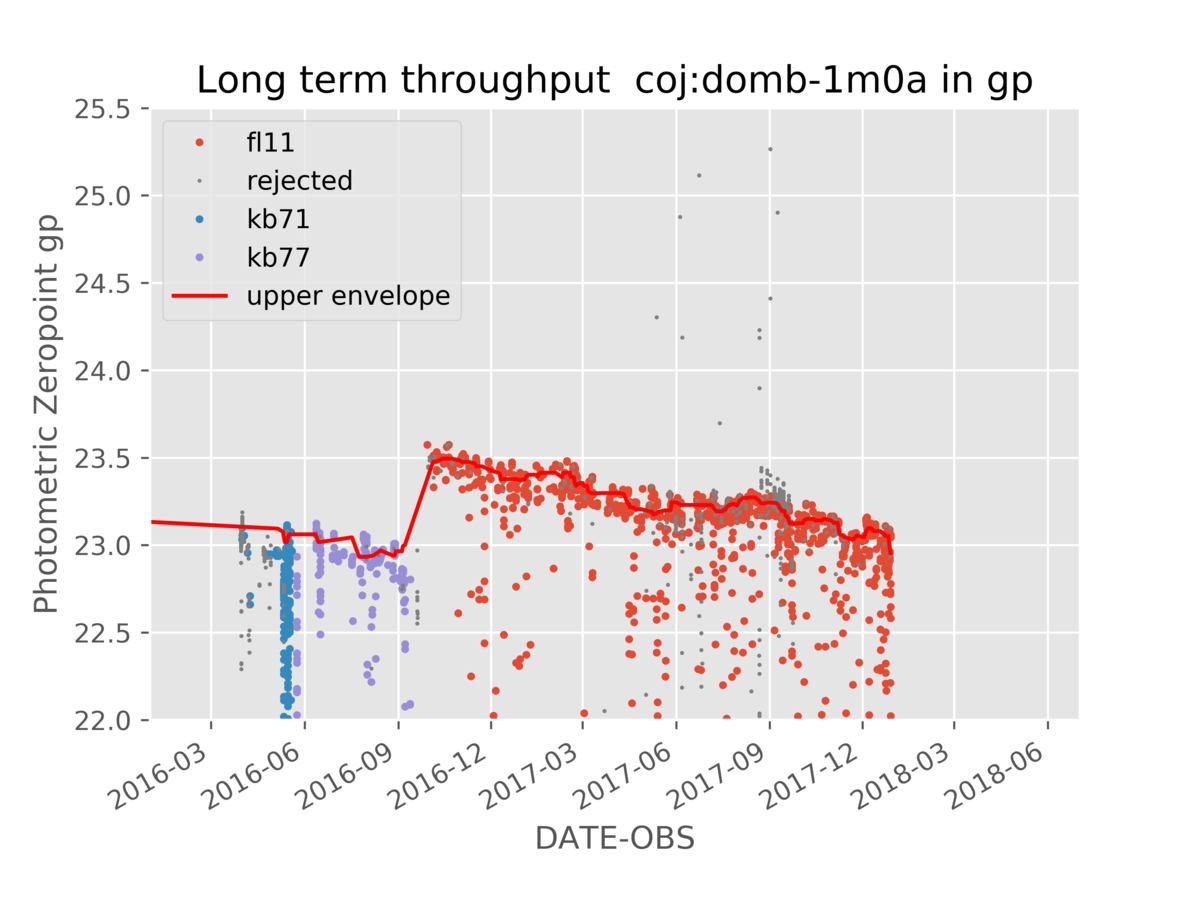} \hspace*{\fill} 
\includegraphics[width=0.49\textwidth]{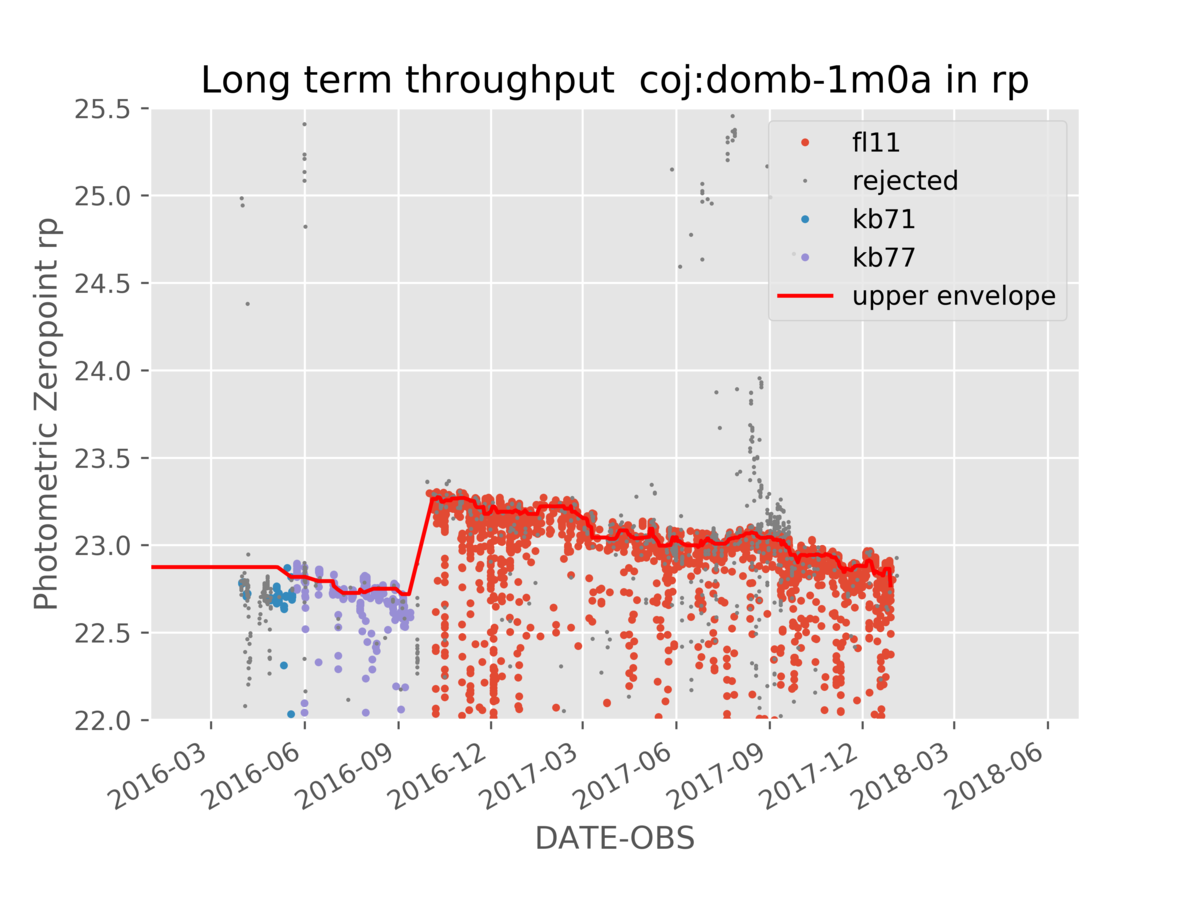} \\[1ex]

OGG  \\ 
\rule{\textwidth}{0.4pt} \\
\includegraphics[width=0.49\textwidth]{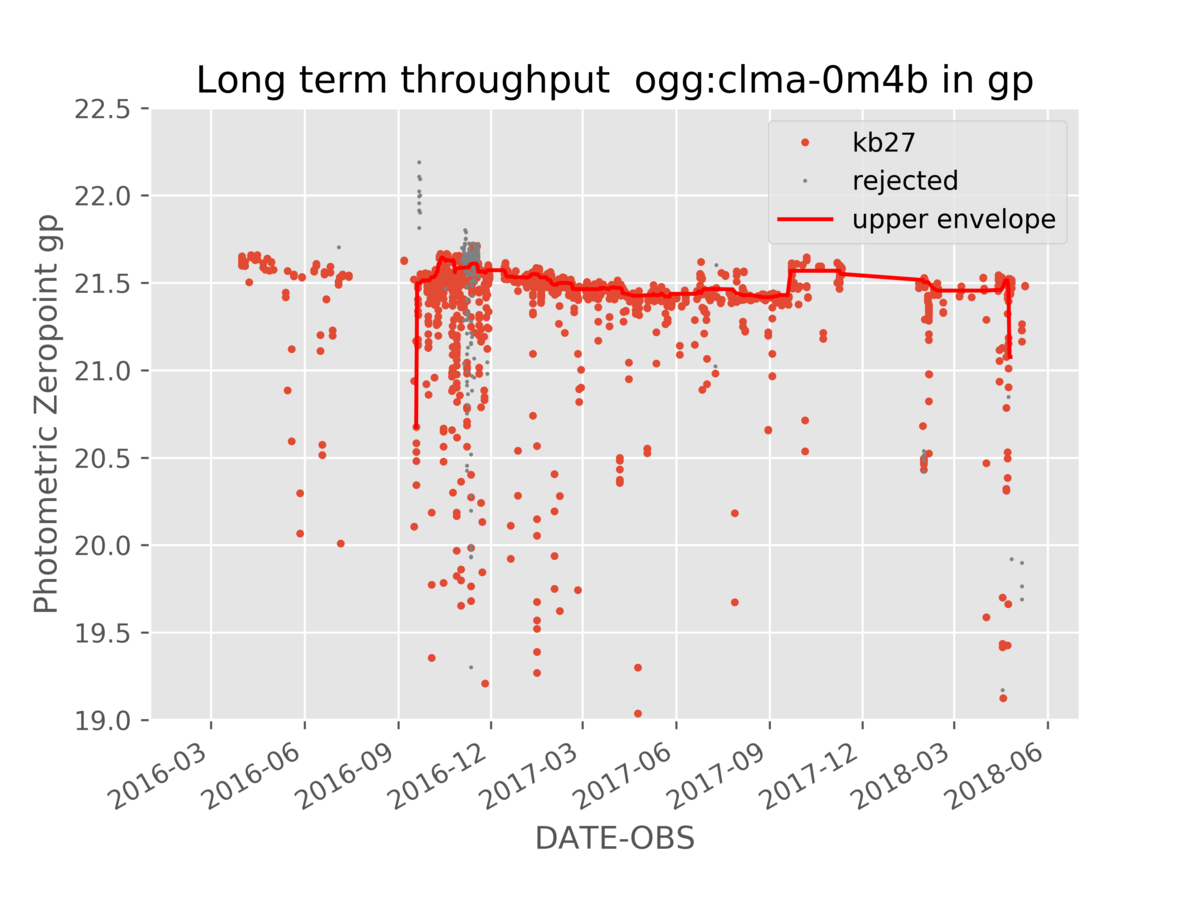} \hspace*{\fill}
\includegraphics[width=0.49\textwidth]{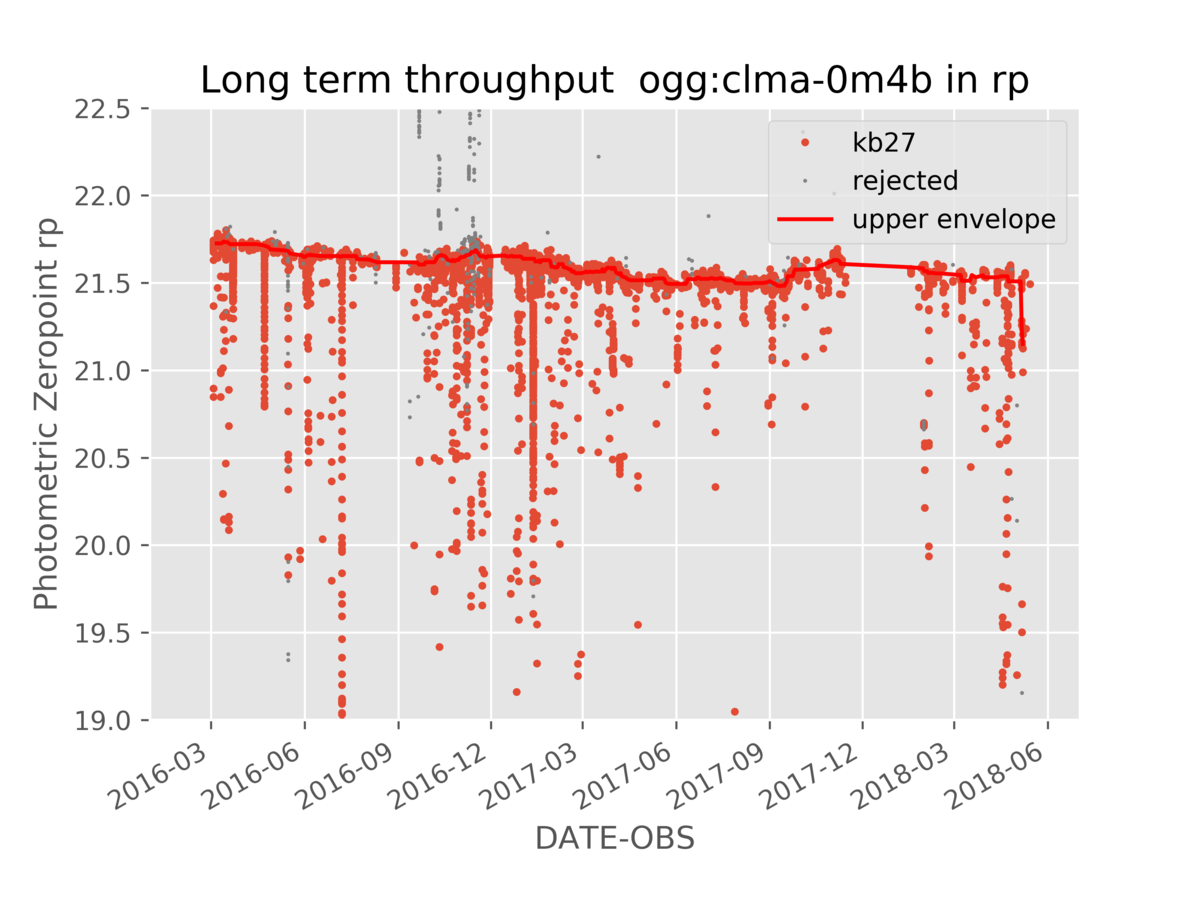} \\
\includegraphics[width=0.49\textwidth]{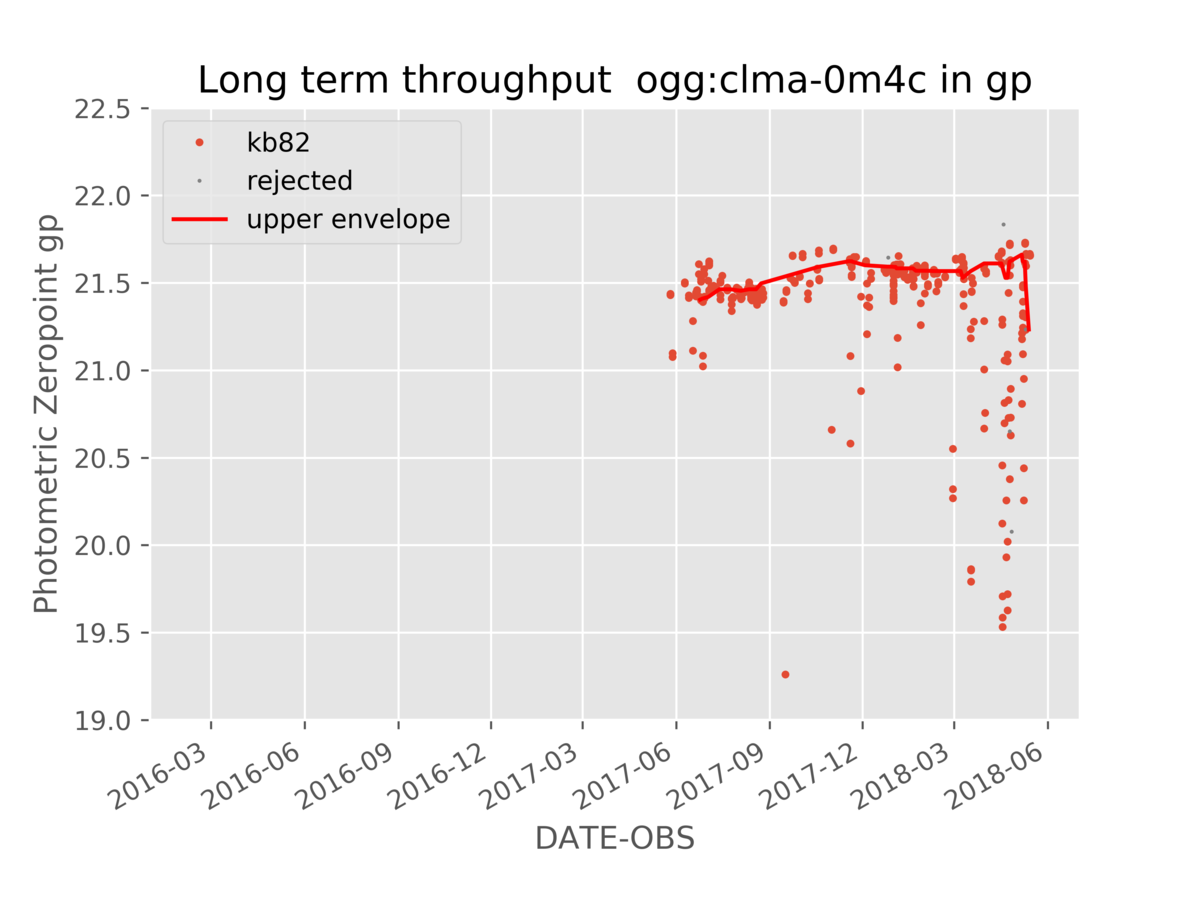} \hspace*{\fill}
\includegraphics[width=0.49\textwidth]{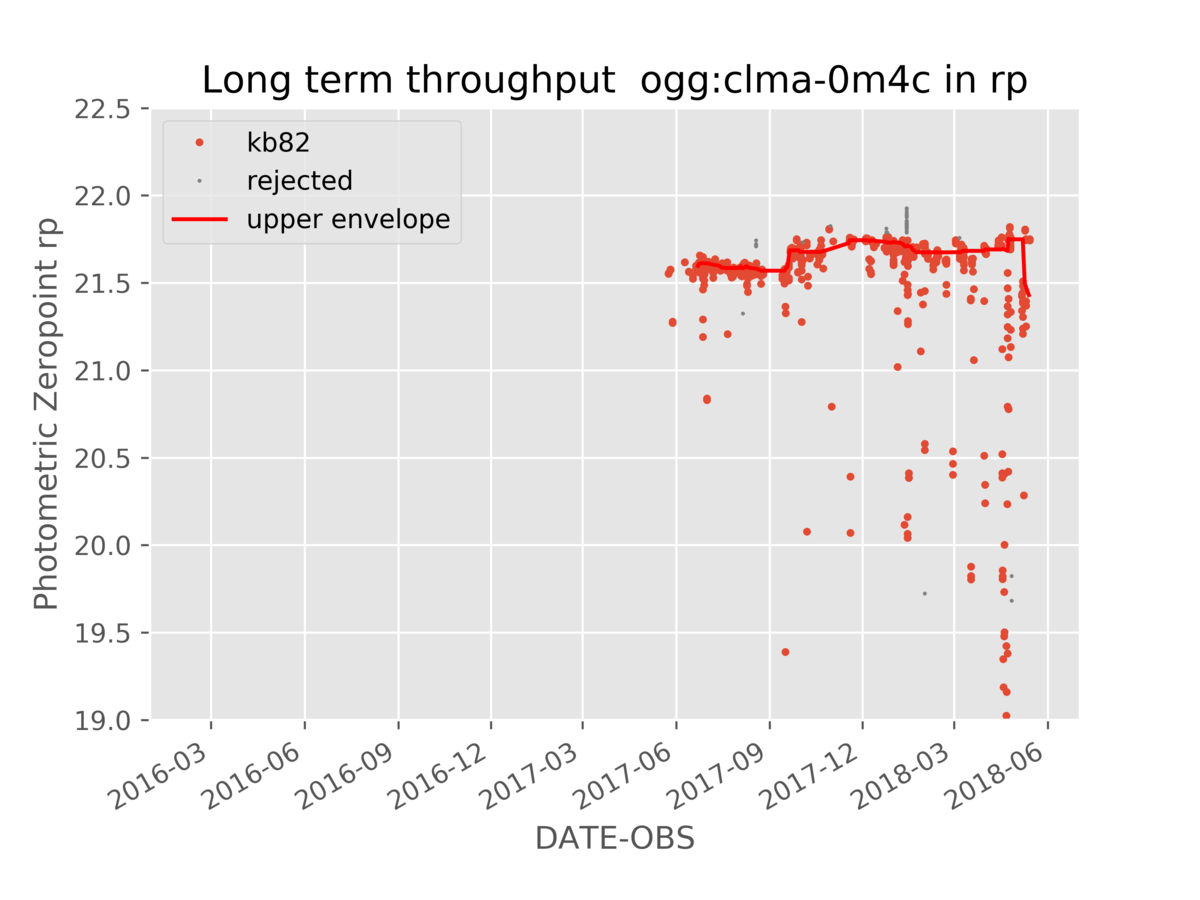} \\[1ex]
\caption {continued}
\end{figure}

\begin{figure}\ContinuedFloat
\centering 
OGG continued \\ 
\rule{\textwidth}{0.4pt} \\
\includegraphics[width=0.49\textwidth]{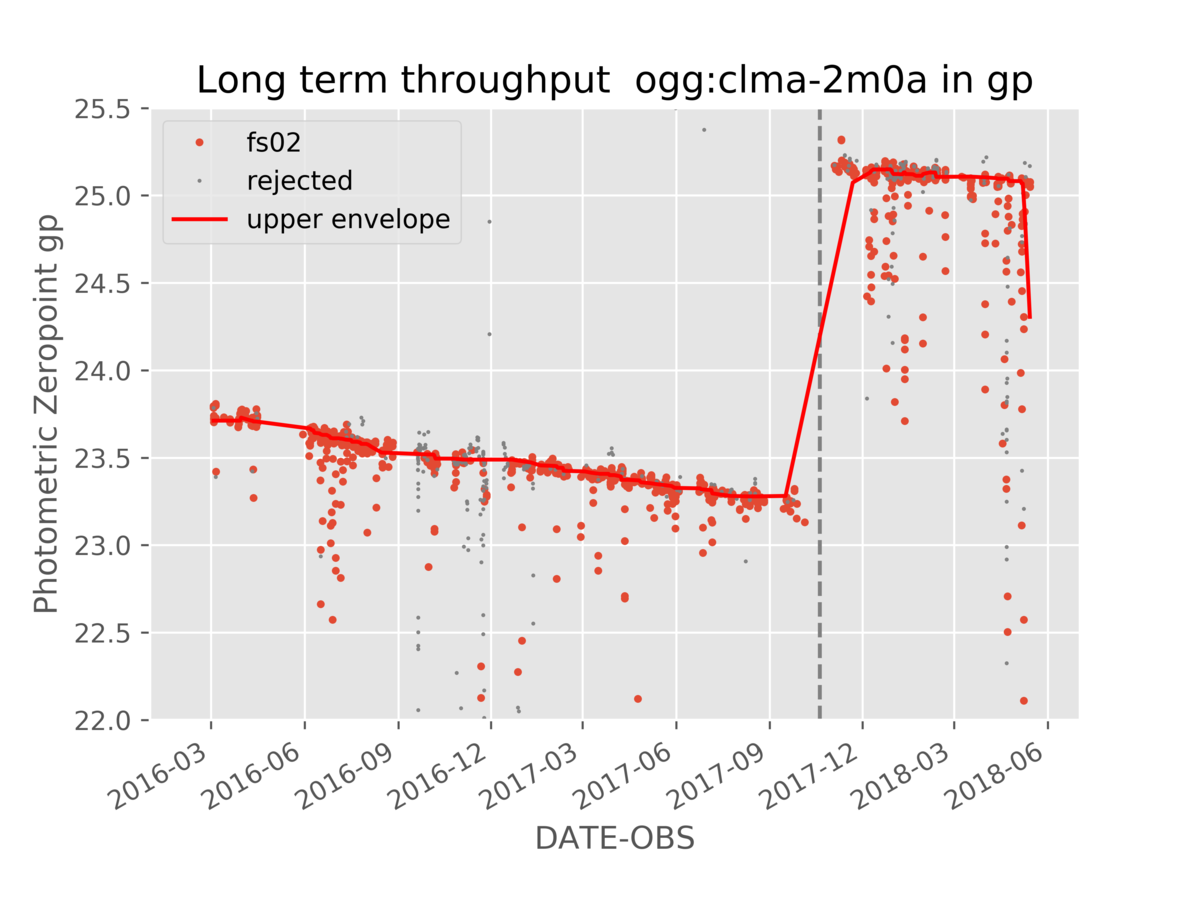} \hspace*{\fill}
\includegraphics[width=0.49\textwidth]{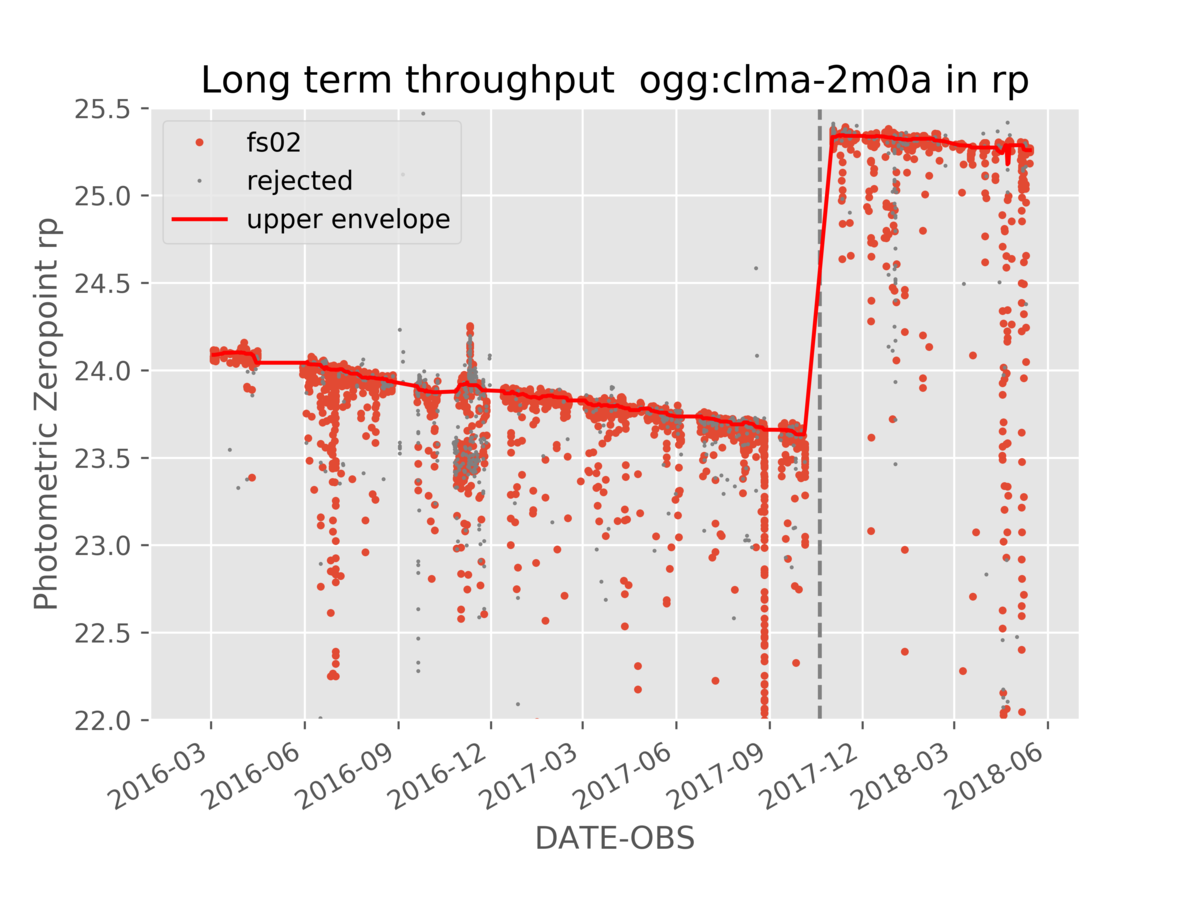} \\[1ex]

ELP  \\ 
\rule{\textwidth}{0.4pt} \\
\includegraphics[width=0.49\textwidth]{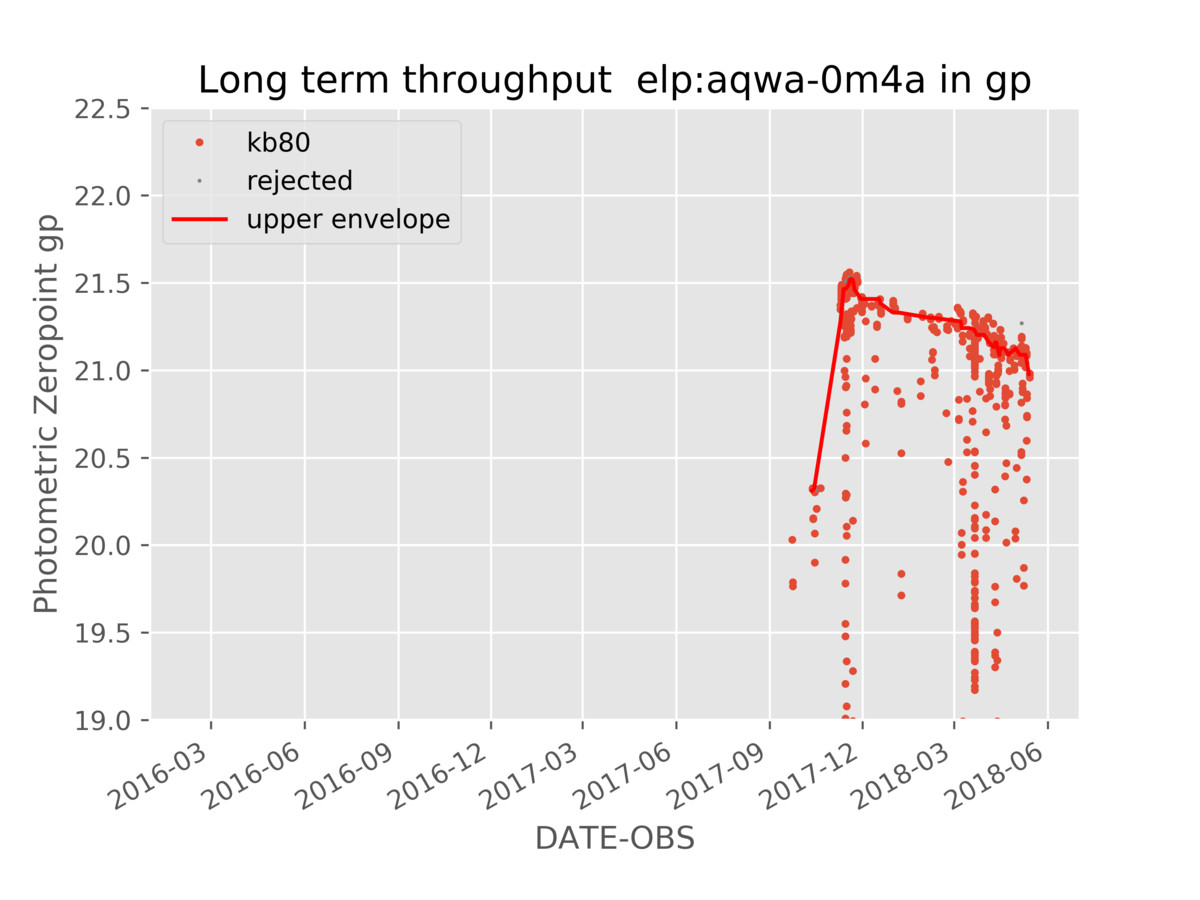} \hspace*{\fill}
\includegraphics[width=0.49\textwidth]{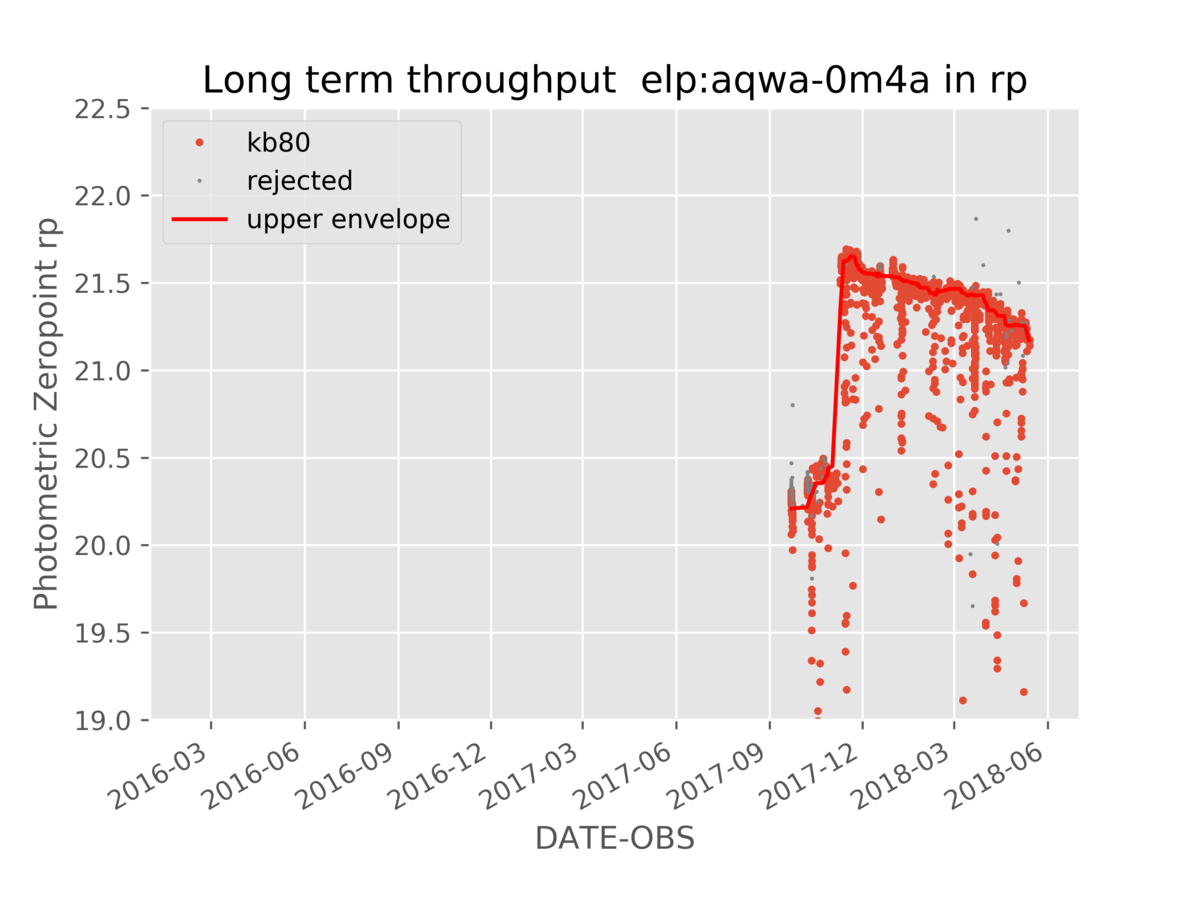} \\
\includegraphics[width=0.49\textwidth]{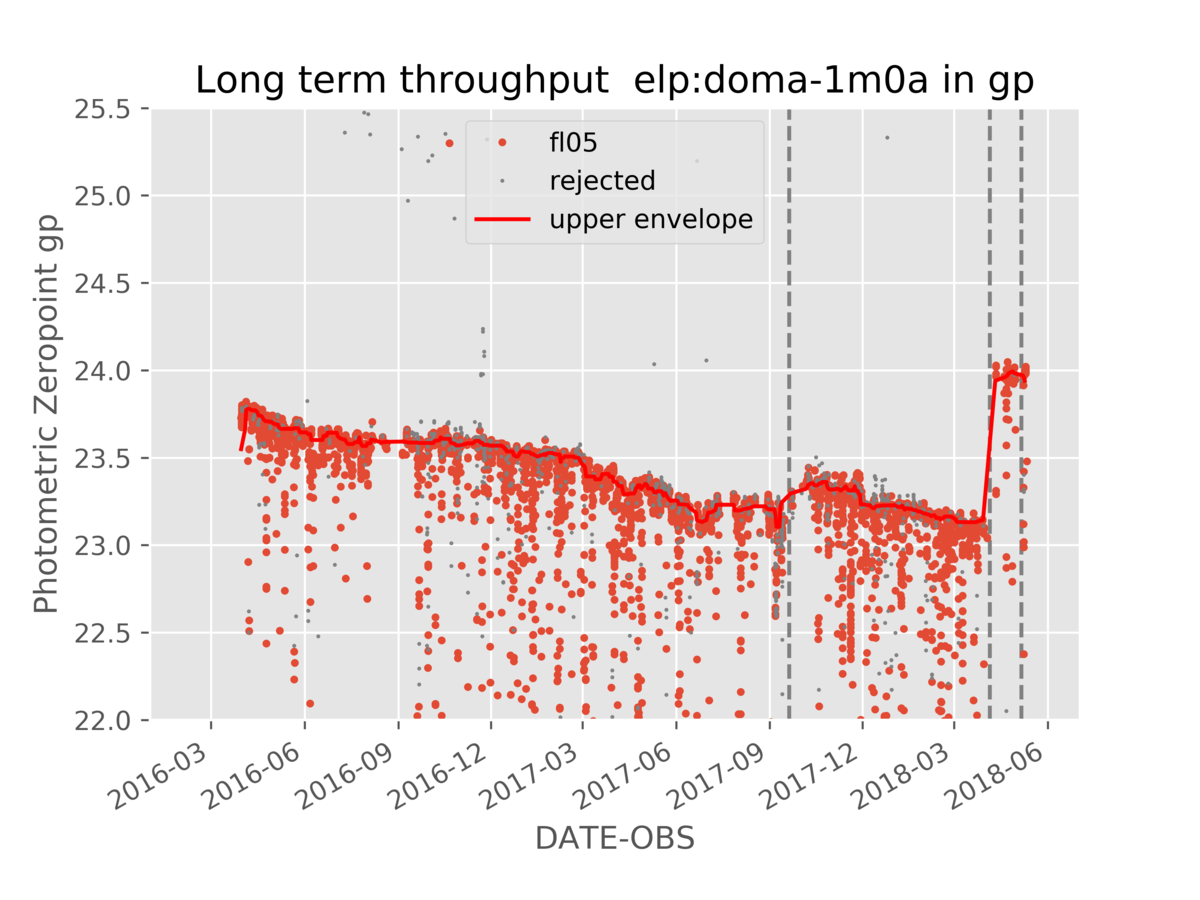} \hspace*{\fill}
\includegraphics[width=0.49\textwidth]{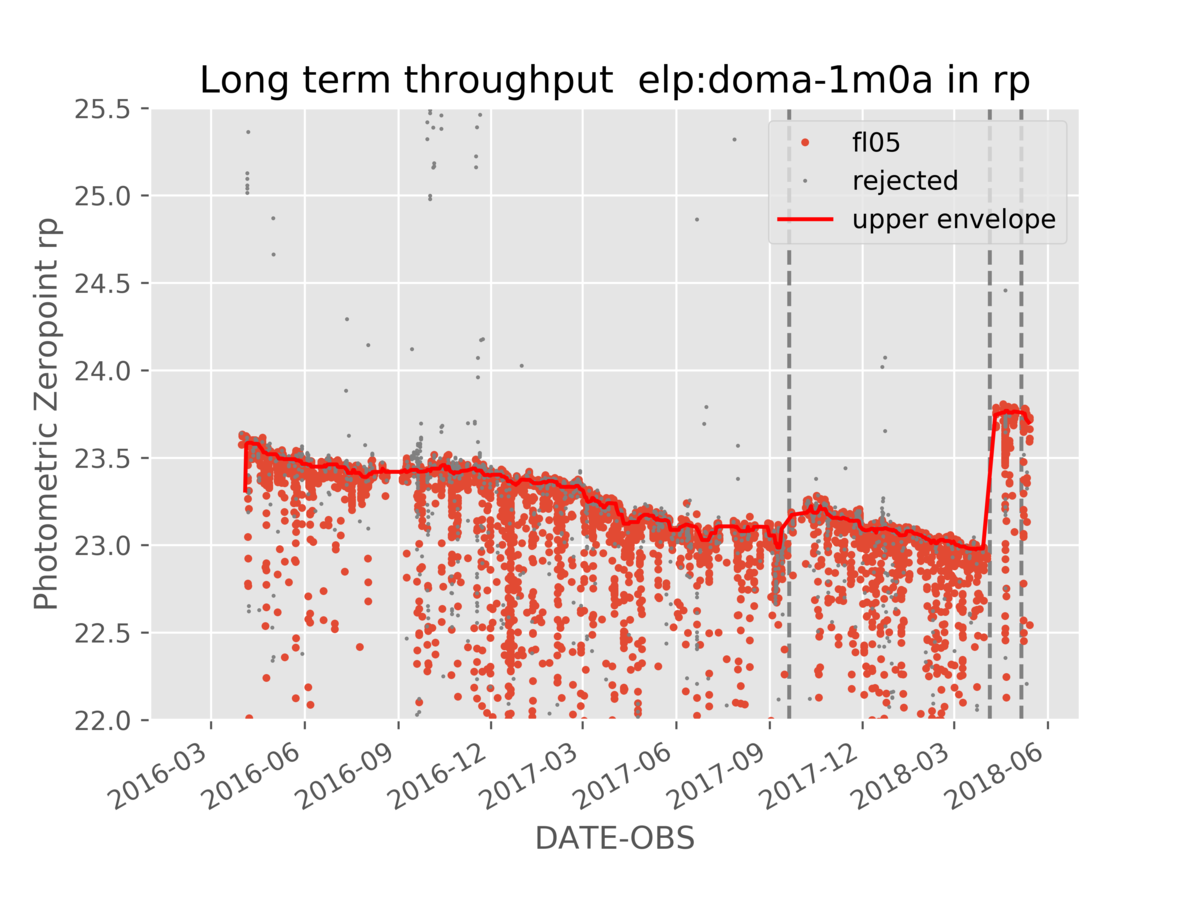} \\[1ex]
\caption {continued}
\end{figure}

\begin{figure}\ContinuedFloat
\centering
CPT \\ 
\rule{\textwidth}{0.4pt} \\
\includegraphics[width=0.49\textwidth]{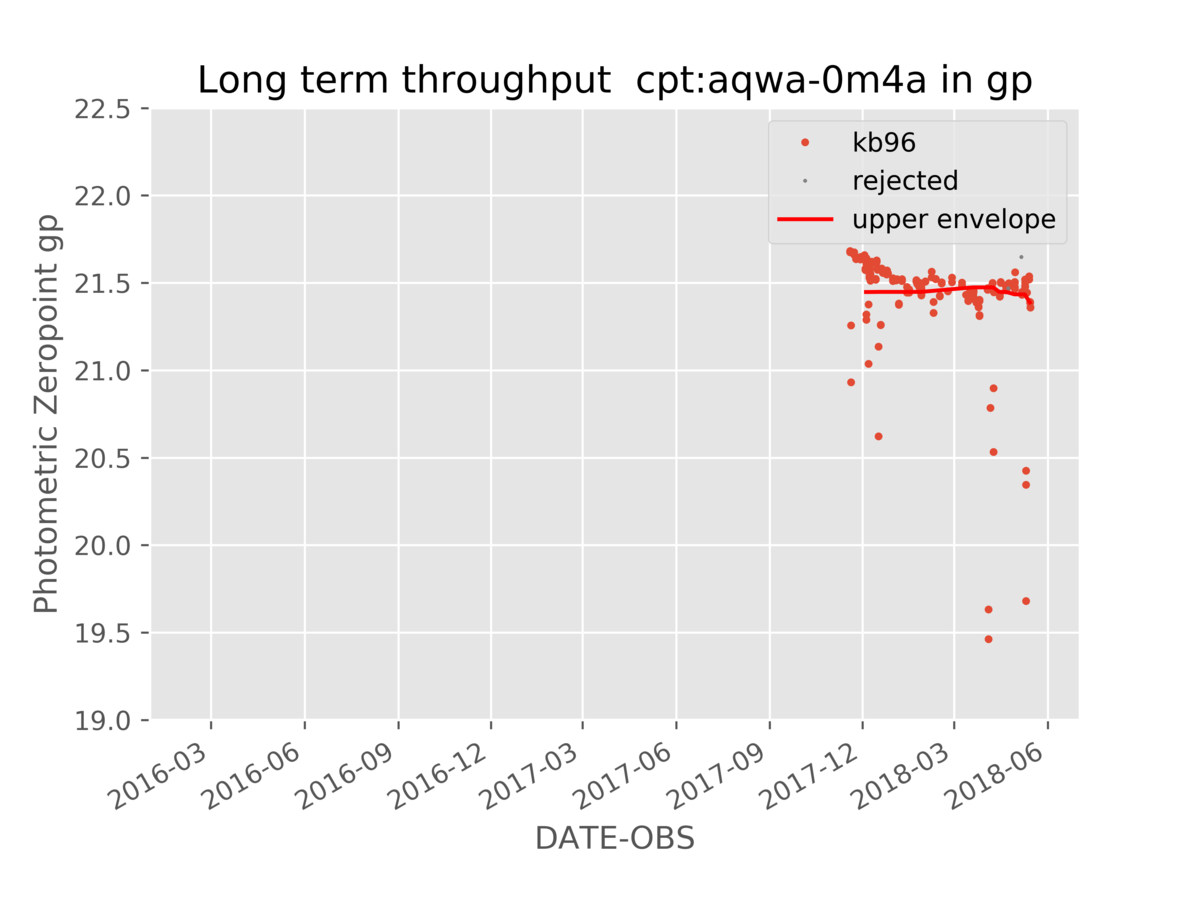} \hspace*{\fill}
\includegraphics[width=0.49\textwidth]{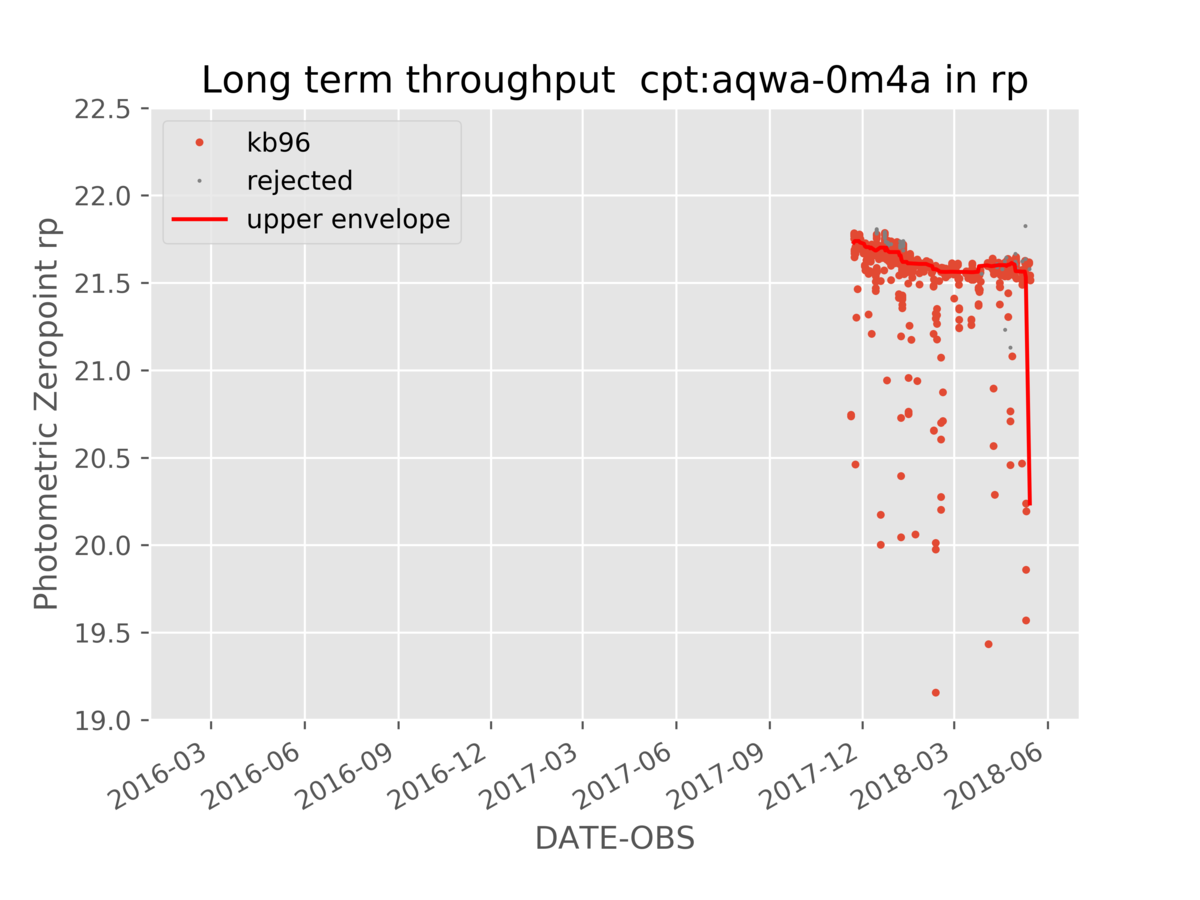} \\
\includegraphics[width=0.49\textwidth]{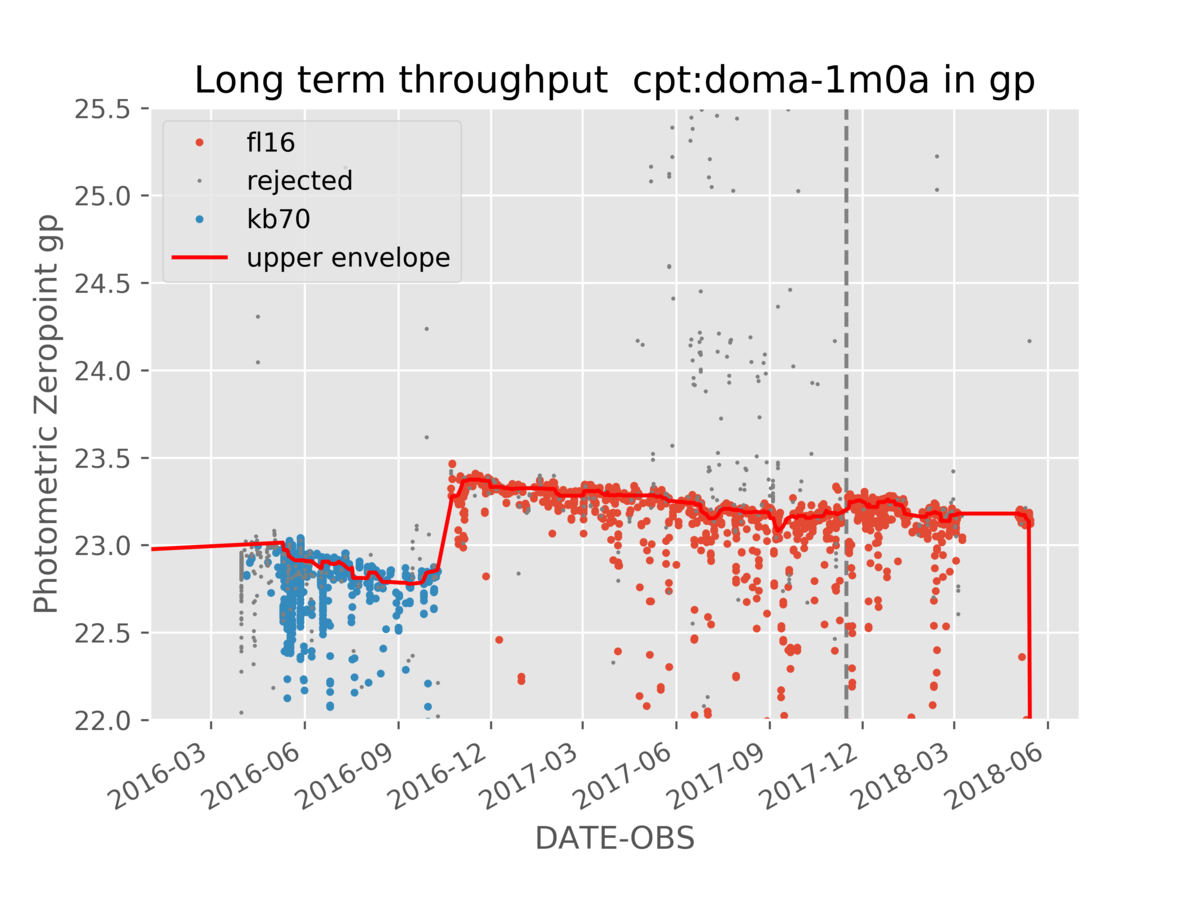} \hspace*{\fill} 
\includegraphics[width=0.49\textwidth]{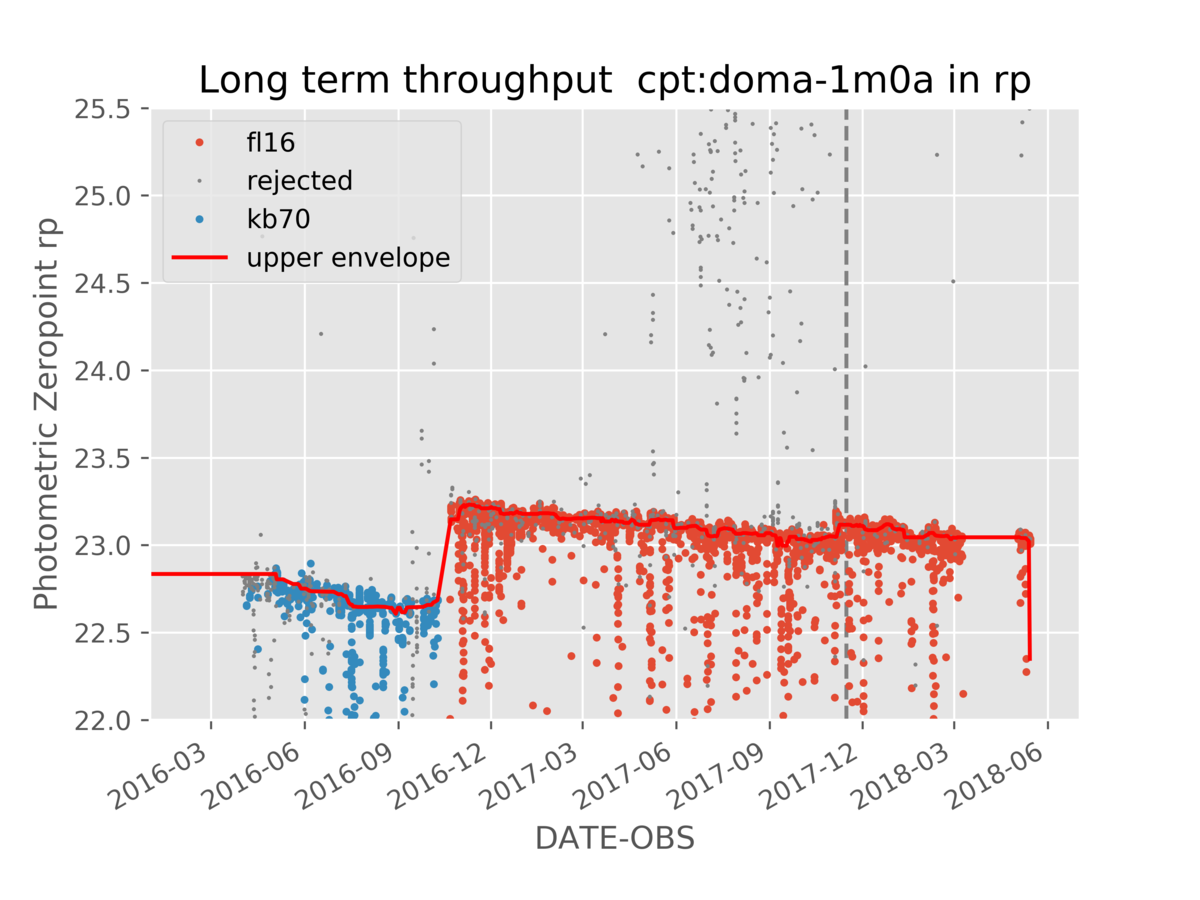} \\
\includegraphics[width=0.49\textwidth]{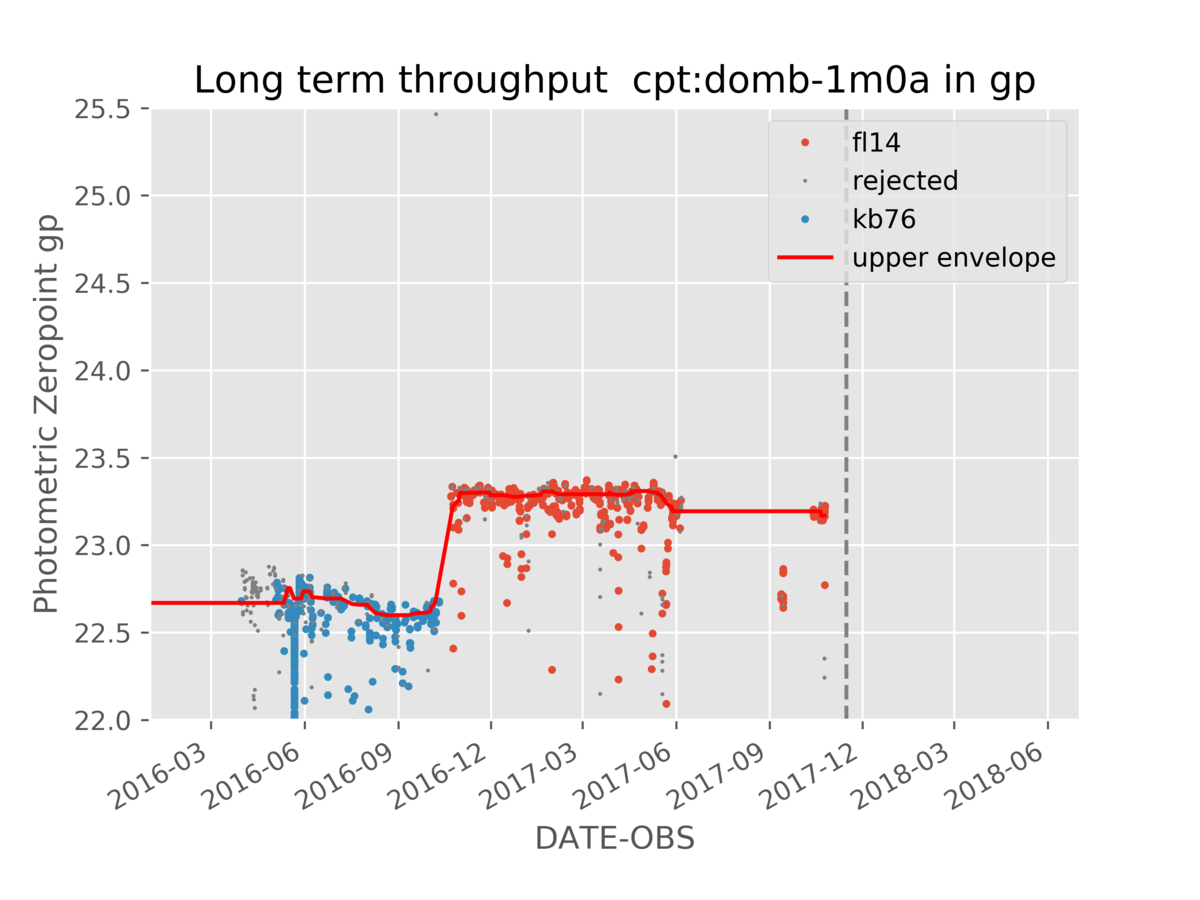} \hspace*{\fill} 
\includegraphics[width=0.49\textwidth]{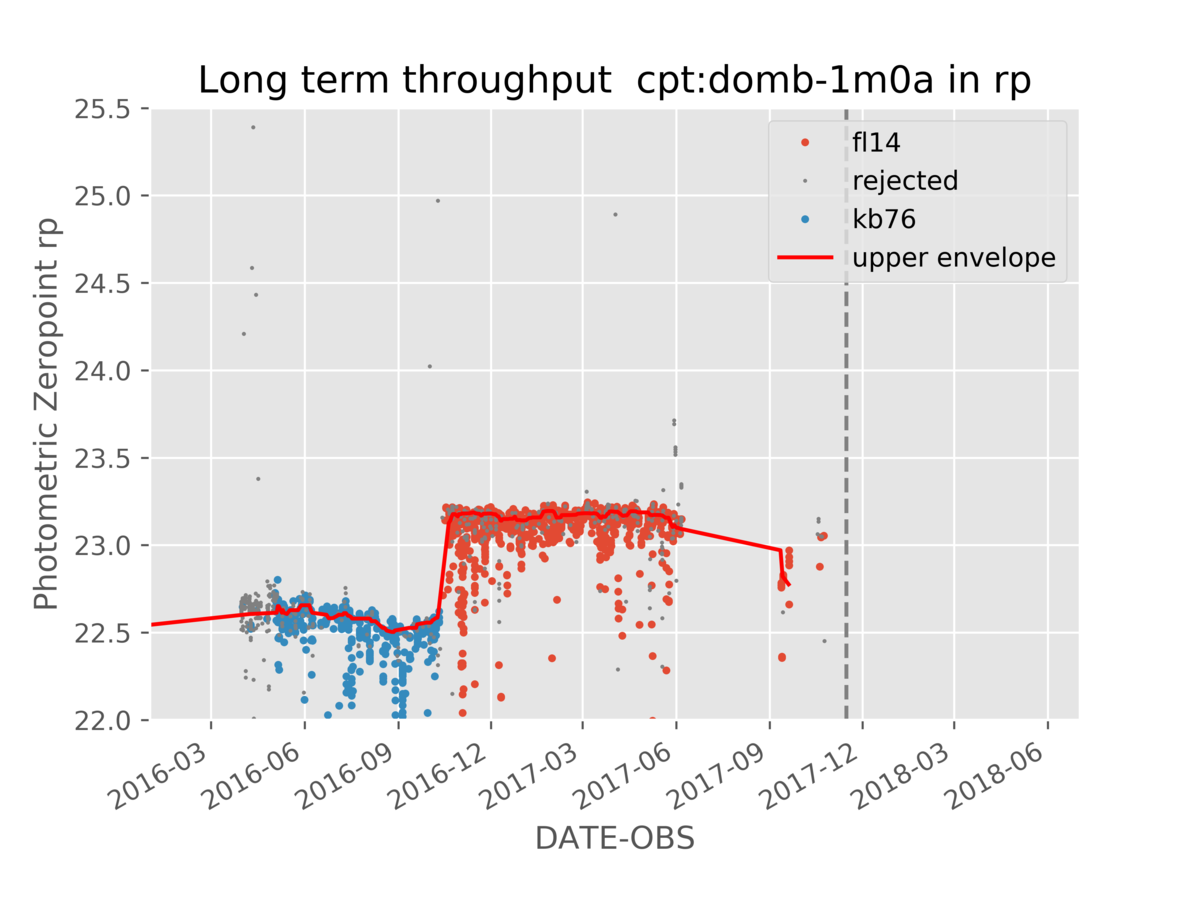} \\[1ex]
\caption {continued}
\end{figure}

\begin{figure}\ContinuedFloat
\centering
CPT continued \\ 
\rule{\textwidth}{0.4pt} \\
\includegraphics[width=0.49\textwidth]{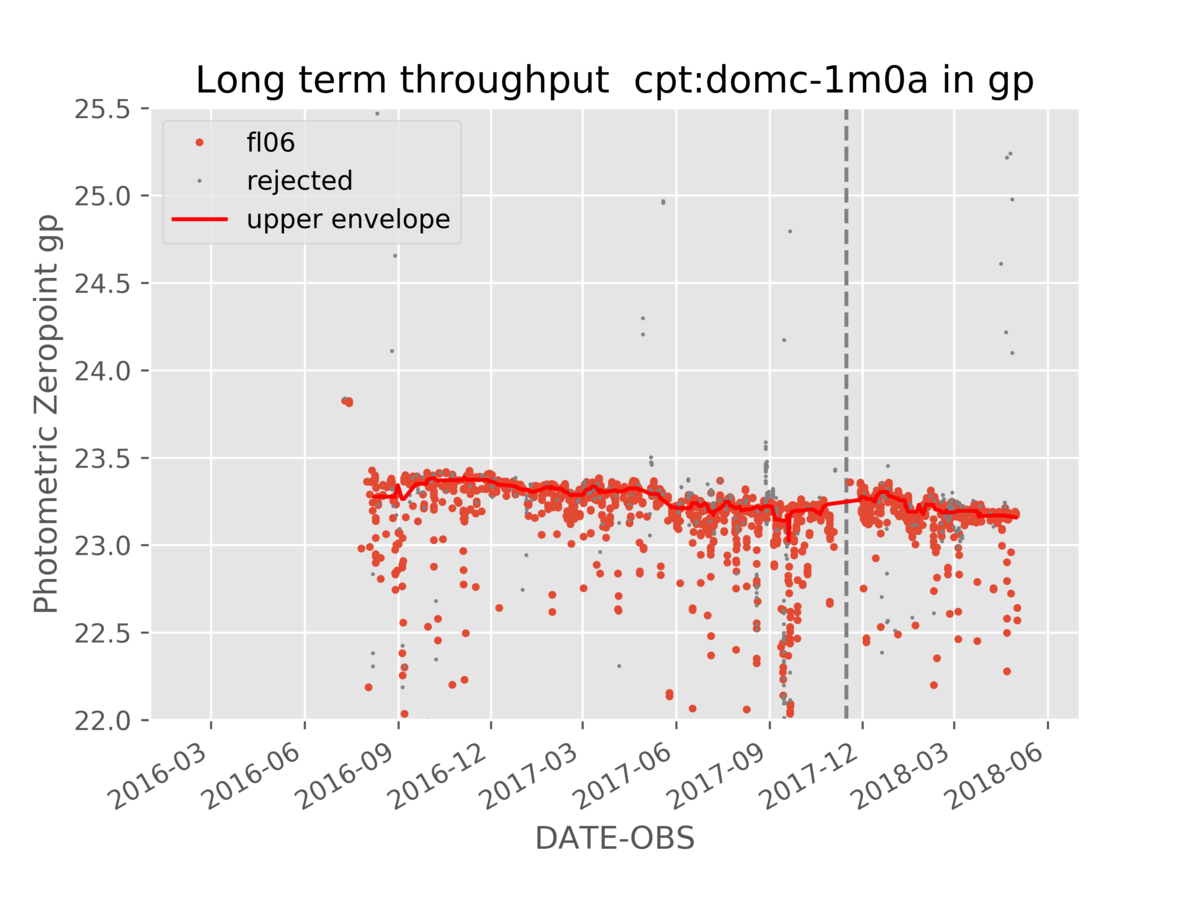} \hspace*{\fill} 
\includegraphics[width=0.49\textwidth]{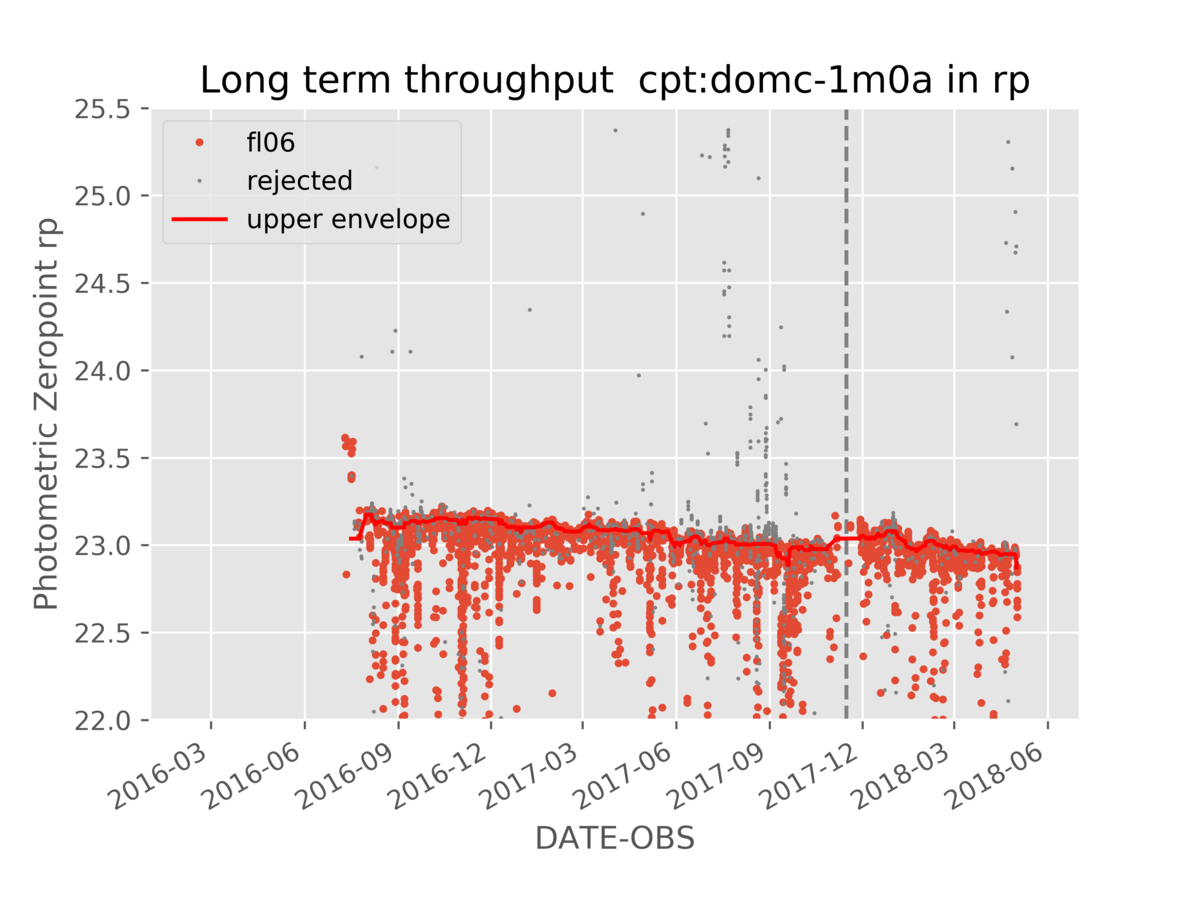} \\[1ex]
TFN \\ 
\rule{\textwidth}{0.4pt} \\
\includegraphics[width=0.49\textwidth]{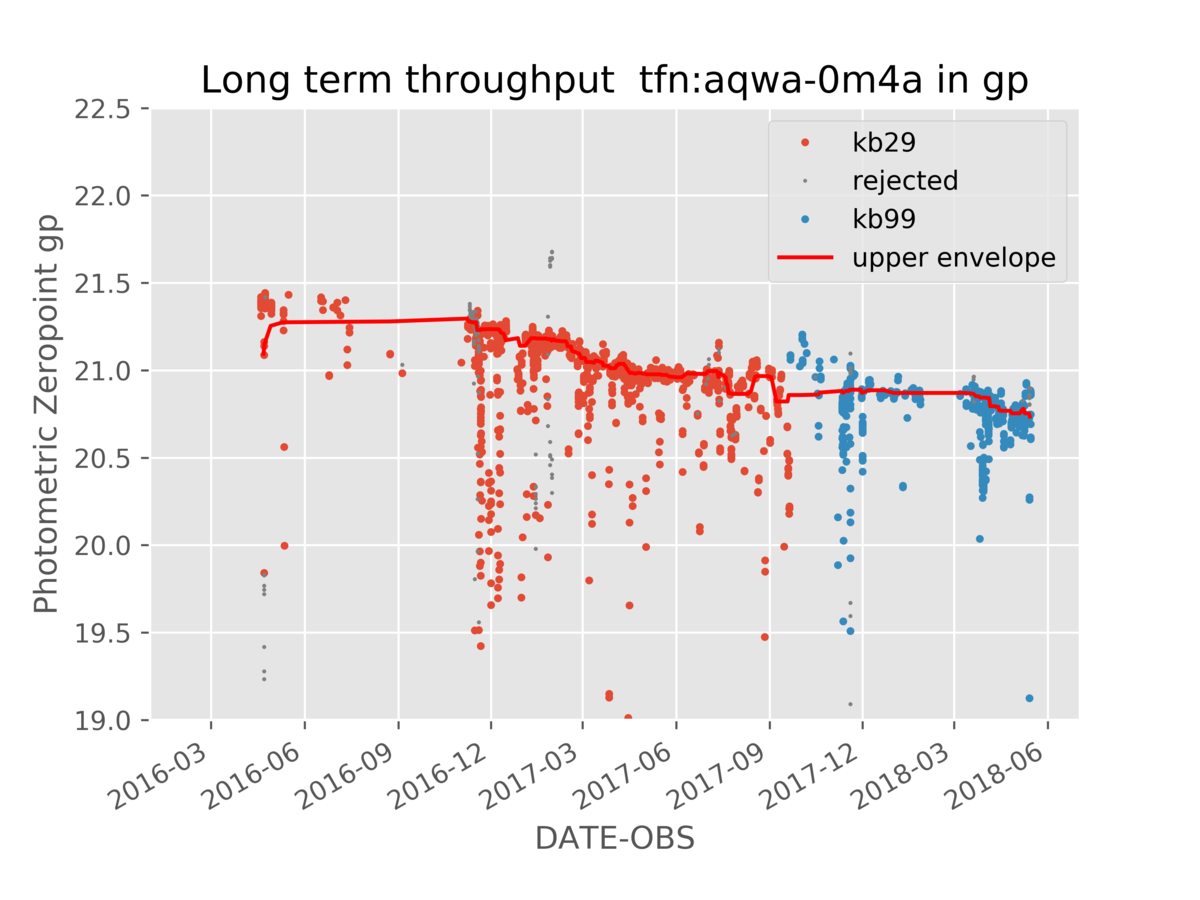} \hspace*{\fill}
\includegraphics[width=0.49\textwidth]{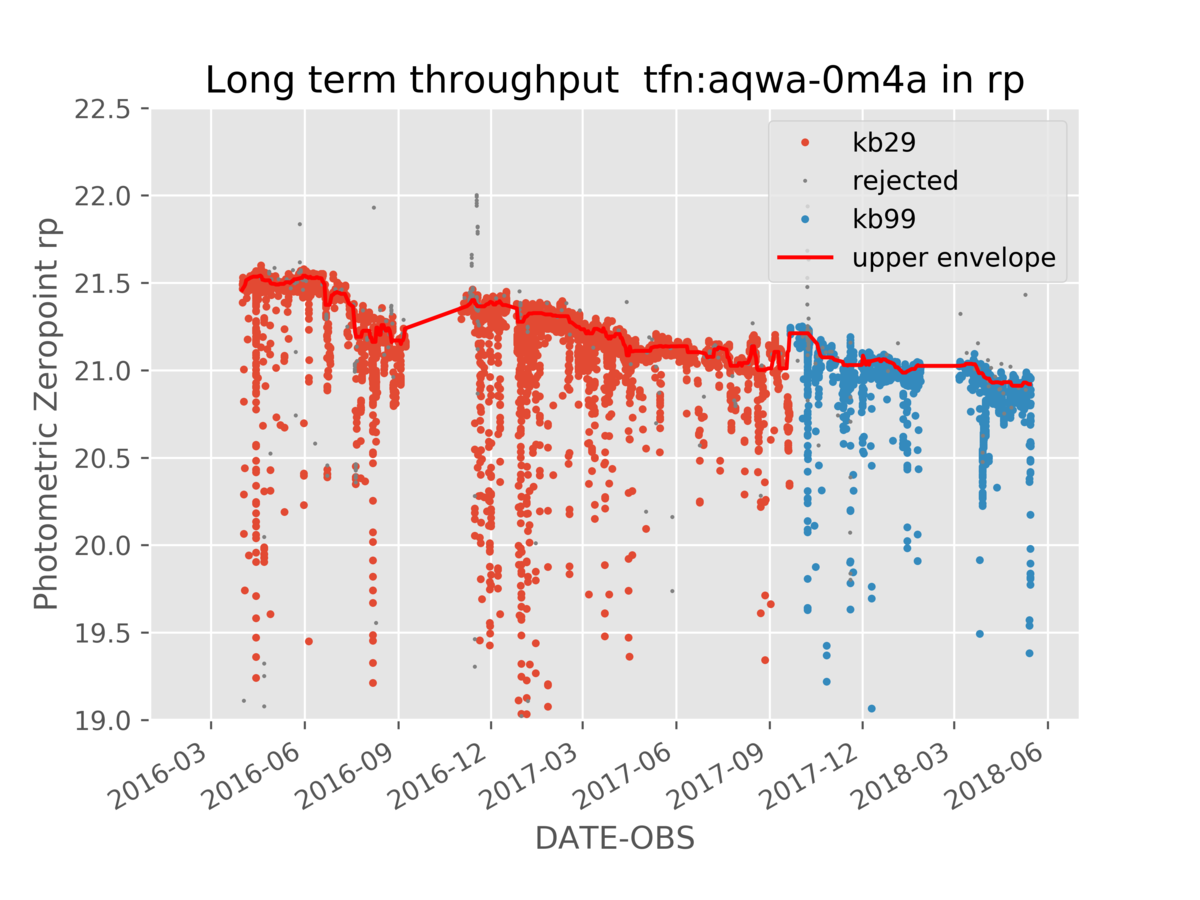} \\
\includegraphics[width=0.49\textwidth]{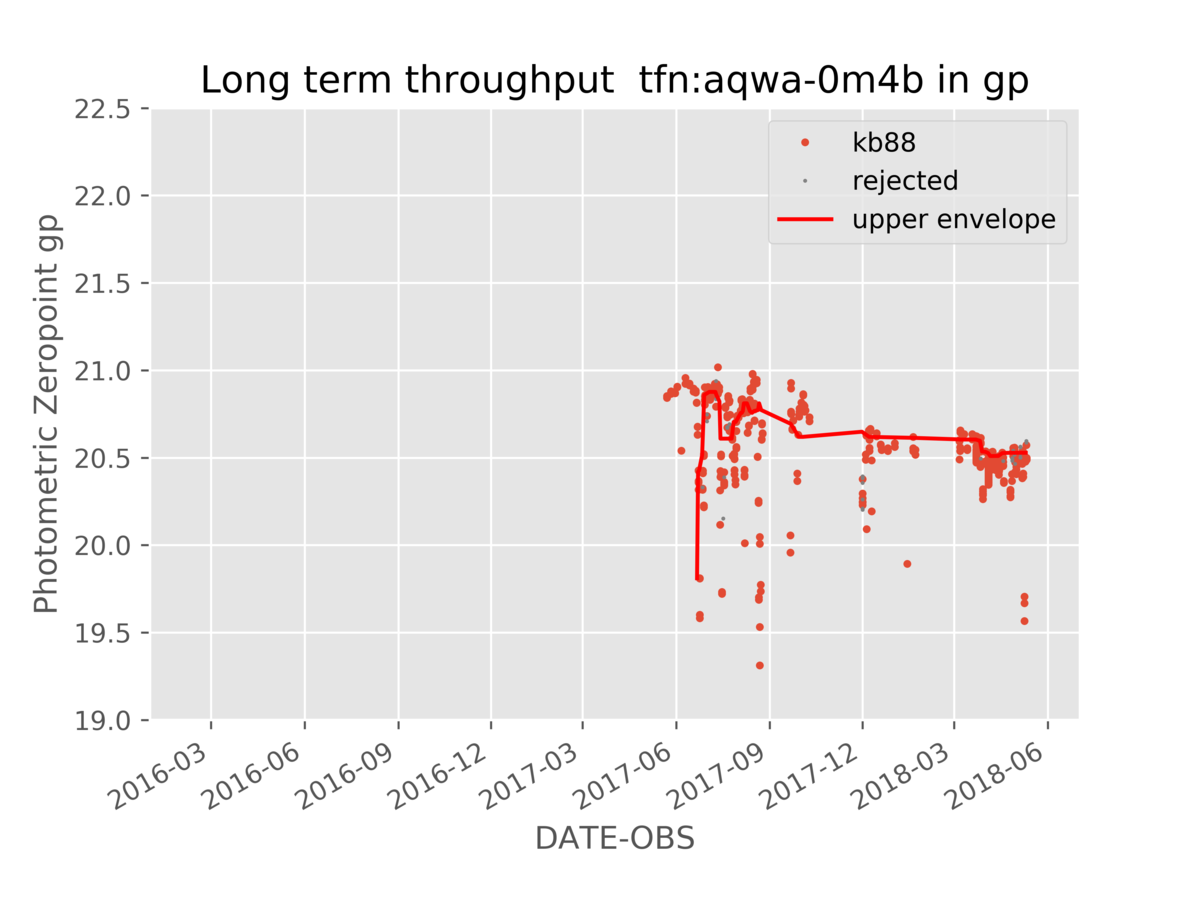} \hspace*{\fill}
\includegraphics[width=0.49\textwidth]{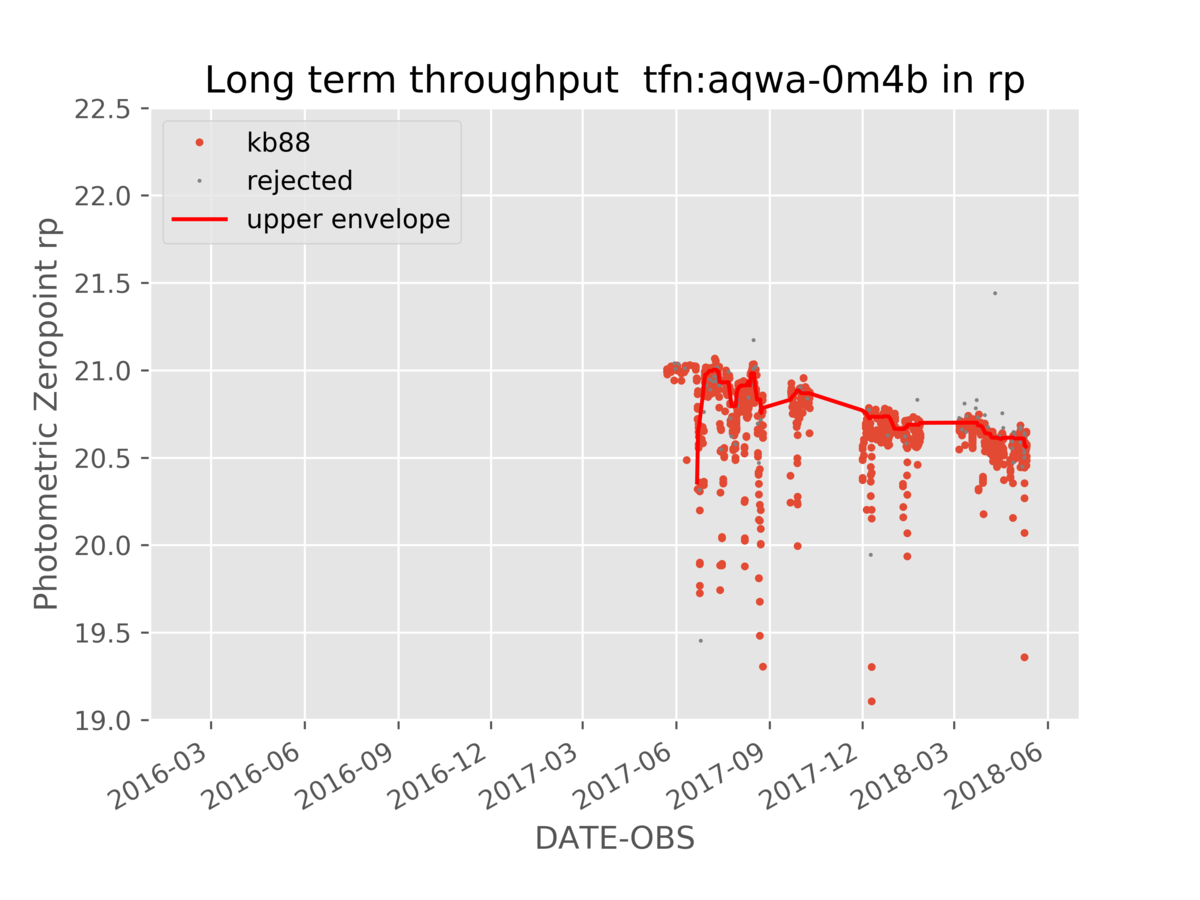} \\[1ex]
\caption {continued}
\end{figure}

\begin{figure}
\includegraphics[width=\textwidth]{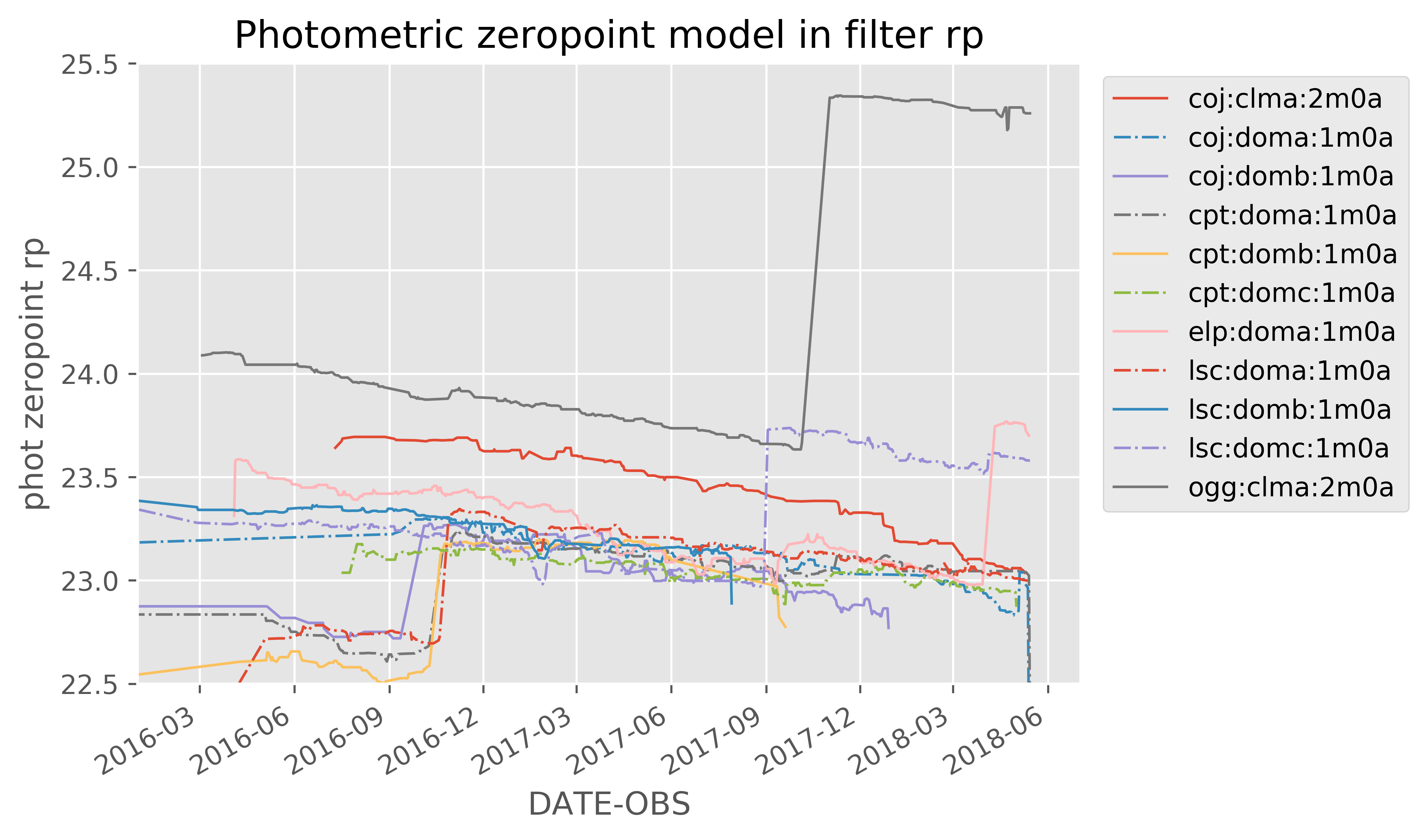}
\caption{\label{fig_alltelescopes} Zeropoint models of all LCOGT 2 meter and 1 meter telescopes combined. }
\end{figure}

We are interested in the behavior of the telescope fleet. We model the upper
envelope of the zeropoint vs. time trend for each individual telescope, as indicated by the red
lines in Figure \ref{fig_zpLSC}. The general trends in the r' band for all 1 meter and 2 meter telescopes
are shown together in Figure \ref{fig_alltelescopes}. We observe:

\begin{enumerate}
\item All telescopes in LCOGT's telescope network lost significant throughput over the time covered
 by the data. 

\item Around October 2016, a number of new imaging cameras (Sinistro) were deployed to the 
telescope, improving the throughput throughout the network significantly.

\item The degradation of the 2 meter telescope mirrors is very significant, especially when seen in
comparison to the performance of the 1 meter telescopes. The recent recoating of the first 2 meter
and 1 meter telescope mirrors demonstrates their  performance potential. Comparing the peak 
performance of the COJ 2 meter telescope with the peak of the ELP 1 meter telescope, as indicated 
in Figure \ref{fig_zpLSC}, yields a  relative throughput difference  ($10^{-0.4\times (25.3-23.8)} 
\sim 3.98$ , which is remarkably close to the expected value based on their aperture ratio 
$(2m)^2/(1m)^2 = 4$.

\item The mirrors at the various sites degrade at a very similar rate, which is remarkable given the
different levels of dust exposures at the sites, e.g., between McDonald observatory and CTIO. The
throughput is lost at a rate of approximately 0.25 mag per year, which corresponds to about $2 \%$
per month, which is significantly higher than the decrease rate of $0.5\%$ reported for the William
Hershel Telescope\cite{designolt,benn2000} for particle deposition on mirrors. This elevated decay
rate is  consistent with the decomposition of the aluminum coating seen throughout our mirrors.

\item We note that since the beginning of 2018, the throughput at the 2 meter telescope at COJ 
(Siding Spring) has started to degrade at an accelerated pace, indicating the approach of a total 
coating failure. It is scheduled for recoating in June 2018.

\item The  mirror recoated in October 2017 at the LSC site has already lost reflectivity of the
 order of 0.2 mag; CO$_2$ snow cleaning in May 2018 (Figure \ref{fig_zpLSC}) did not fully recover
 the throughput of the freshly replaced mirror,  indicating that the current cleaning schedule or
 method is insufficient to maintain a high throughput over an extended time.

\end{enumerate}

\section{Summary and Conclusion}

The vast data set from a large fleet of telescopes with identical instrumentation over a multi-year
time span has provided a unique insight into the aging of the optical surfaces of telescopes at
multiple sites across the globe. At LCOGT, quantifying the throughput evolution as shown in this
paper was an important tool to communicate the urgency of a systematic approach to maintaining the
mirrors and proved more effective than isolated reports about mirrors deteriorating.

LCOGT is evolving from the construction phase to an observatory in full operation; at this
transition point, a new culture is developing to emphasize maintenance and protecting aging equipment
against decay. The significance of the throughput loss at the telescopes was recognized in 2017, and
a process was started to recoat all mirrors within LCOGT's network.  This was especially true once
the drastic improvements from the first three refurbished systems became visible.  Analyzing the
state of the telescopes in near real-time will inform staff in the cost-effective scheduling of
cleaning and recoating of mirrors should telescopes degrade faster than expected (e.g., due to dust
storms).

Understanding the telescope performance reaches beyond managing operational costs: LCOGT operates a
network of telescopes. When an observing request is entered into the system, it is not yet clear at
which telescope it will be executed. Moreover, due to the spread of locations, one of LCOGT's 
unique capabilities is to monitor an object continuously, handing over observations from one site  
to another as the darkness progresses over the globe. In this operation scenario it is problematic
if the throughput varies significantly between telescopes of a class, which are all treated equally 
by the scheduler.  It is hence desirable to maintain the telescopes in an equal {\it and} good
state, or alternatively give the scheduler awareness of the individual telescopes' performance.
Deriving the throughput of all telescopes, from all science images, and on a daily basis can inform 
observatory staff to better the observatory hardware and software.

\section{Acknowledgement}
The Pan-STARRS1 Surveys (PS1) and the PS1 public science archive have been made possible through
contributions by the Institute for Astronomy, the University of Hawaii, the Pan-STARRS Project
Office, the Max-Planck Society and its participating institutes, the Max Planck Institute for
Astronomy, Heidelberg and the Max Planck Institute for Extraterrestrial Physics, Garching, The Johns
Hopkins University, Durham University, the University of Edinburgh, the Queen's University Belfast,
the Harvard-Smithsonian Center for Astrophysics, the Las Cumbres Observatory Global Telescope
Network Incorporated, the National Central University of Taiwan, the Space Telescope Science
Institute, the National Aeronautics and Space Administration under Grant No. NNX08AR22G issued
through the Planetary Science Division of the NASA Science Mission Directorate, the National Science
Foundation Grant No. AST-1238877, the University of Maryland, Eotvos Lorand University (ELTE), the
Los Alamos National Laboratory, and the Gordon and Betty Moore Foundation.

\bibliography{bib}
\bibliographystyle{spiejour}

\end{document}